\documentclass[prl,aps,amssymb,nofootinbib,floatfix,reprint,notitlepage]{revtex4-1} 

\usepackage[dvipsnames]{xcolor}
\usepackage{amsmath,amssymb,bbold,bm}
\usepackage{graphicx}
\usepackage[normalem]{ulem}
\usepackage{cancel}
\usepackage{comment}

\usepackage[pdfpagelabels,colorlinks=true,breaklinks=true,linktocpage=true,linkcolor=Blue,urlcolor=Blue,citecolor=Blue,%
pdftitle={Emergent Spinon Dispersion and Symmetry Breaking in Two-channel Kondo Lattices},%
pdfauthor={Yang Ge, Yashar Komijani}%
]{hyperref}

\newcommand{\sbraket}[1]{\langle #1 \rangle}
\newcommand{\be}{\begin{equation}}
\newcommand{\ben}{\begin{equation*}}
\newcommand{\ee}{\end{equation}}
\newcommand{\een}{\end{equation*}}
\newcommand{\bs}{\begin{split}}
\newcommand{\es}{\end{split}}
\newcommand{\bmx}{\begin{array}}
\newcommand{\emx}{\end{array}}
\newcommand{\muad}{\hspace{.1cm}}
\newcommand{\bea}{\begin{eqnarray}}
\newcommand{\bean}{\begin{eqnarray*}}
\newcommand{\eea}{\end{eqnarray}}
\newcommand{\eean}{\end{eqnarray*}}
\newcommand{\dg}{^{\dagger}}
\newcommand{\dn}{^{\vphantom{\dagger}}}

\newcommand{\ua}{\uparrow}
\newcommand{\da}{\downarrow}
\newcommand{\Ua}{\Uparrow}
\newcommand{\Da}{\Downarrow}
\newcommand{\bb}[1]{\mathbb{#1}}

\newcommand{\eps}{\epsilon}

\newcommand{\sgn}[1]{{\rm sign}{#1}}
\newcommand{\pref}[1]{(\ref{#1})}

\newcommand{\intinf}[1]{\int_{-\infty}^{+\infty}{#1}}
\newcommand{\intoinf}[1]{\int_{0}^{\infty}{#1}}

\newcommand{\im}[1]{{\rm Im}\left[ #1 \right]}

\newcommand{\abs}[1]{\left\vert #1 \right\vert}
\newcommand{\bra}[1]{\left\langle #1 \right\vert}
\newcommand{\ket}[1]{\left\vert #1\right\rangle}
\newcommand{\braket}[1]{\left\langle #1\right\rangle}

\newcommand{\sepline}{}

\newcommand{\bw}[1]{\begin{widetext}}
\newcommand{\ew}[1]{\end{widetext}}

\newcommand{\gray}[1]{}

\newcommand{\nothing}[1]{}

\begin{document}

\title{Emergent Spinon Dispersion and Symmetry Breaking in Two-Channel Kondo Lattices}
\author{Yang Ge}
\author{Yashar Komijani\,$^{*}$}
 \affiliation{ Department of Physics, University of Cincinnati, Cincinnati, Ohio 45221, USA}
\date{\today}
\begin{abstract}
Two-channel Kondo lattice serves as a model for a growing family of heavy-fermion compounds. We employ a dynamical large-$N$ technique and go beyond the independent bath approximation to study this model both numerically and analytically using renormalization group ideas. We show that the Kondo effect induces dynamic magnetic correlations that lead to an emergent spinon dispersion.
Furthermore, we develop a quantitative framework that interpolates between infinite dimension where the channel-symmetry broken results of mean-field theory are confirmed, and one-dimension where the channel symmetry is restored and a critical fractionalized mode is found.

\end{abstract}
\maketitle
The screening of a magnetic impurity by the conduction electrons in a metal is governed by the Kondo effect. 
The multichannel version is when several channels compete for 
a single impurity, as a result of which the spin is frustrated and a new critical ground state formed with a fractional residual impurity entropy. In the two-channel case, this entropy $\frac{1}{2}\log 2$ corresponds to a Majorana fermion. If the channel symmetry is broken, the weaker channels decouple and the stronger-coupled channels \emph{win} to screen the impurity at low temperature \cite{Andrei84,Affleck92,Emery1992,Affleck1993}.

While the case of a single impurity is well understood, much less is known about Kondo lattices where a lattice of spins is screened by conduction electrons~\cite{Hewson,Si2014,Coleman2015}, especially if multiple conduction channels are involved~\cite{Cox1996}. The most established fact is the prediction of a large Fermi surface (FS) in the Kondo-dominated regime of the single-channel Kondo lattice~\cite{Oshikawa00}.
In the multichannel case, the continuous channel symmetry naturally leads to new patterns of entanglement which are potentially responsible for the non-Fermi liquid physics \cite{Jarrell96,Jarrell1997}, symmetry breaking, and possibly fractionalized order parameter \cite{ofc}. This partly arises from the fact that the residual entropy seen in the impurity has to eventually disappear at zero temperature in the case of a lattice. 

Beside fundamental interest, a pressing reason for studying this physics is that the multichannel Kondo lattice (MCKL), and in particular 2CKL, seems to be an appropriate model for several heavy-fermion compounds, e.g.\ the family of Pr\textsl{Tr}$_2$Zn$_{20}$ (\textsl{Tr}=Ir,Rh)~\cite{Onimaru2016,Patri2020} as well as recent proposals that MCKLs may support nontrivial topology \cite{Hu21,Rebecca21} and  non-Abelian Kondo anyons~\cite{Lopes20,Komijaniqubit}.

\begin{figure}[tp!]
	\includegraphics[width=1\linewidth]{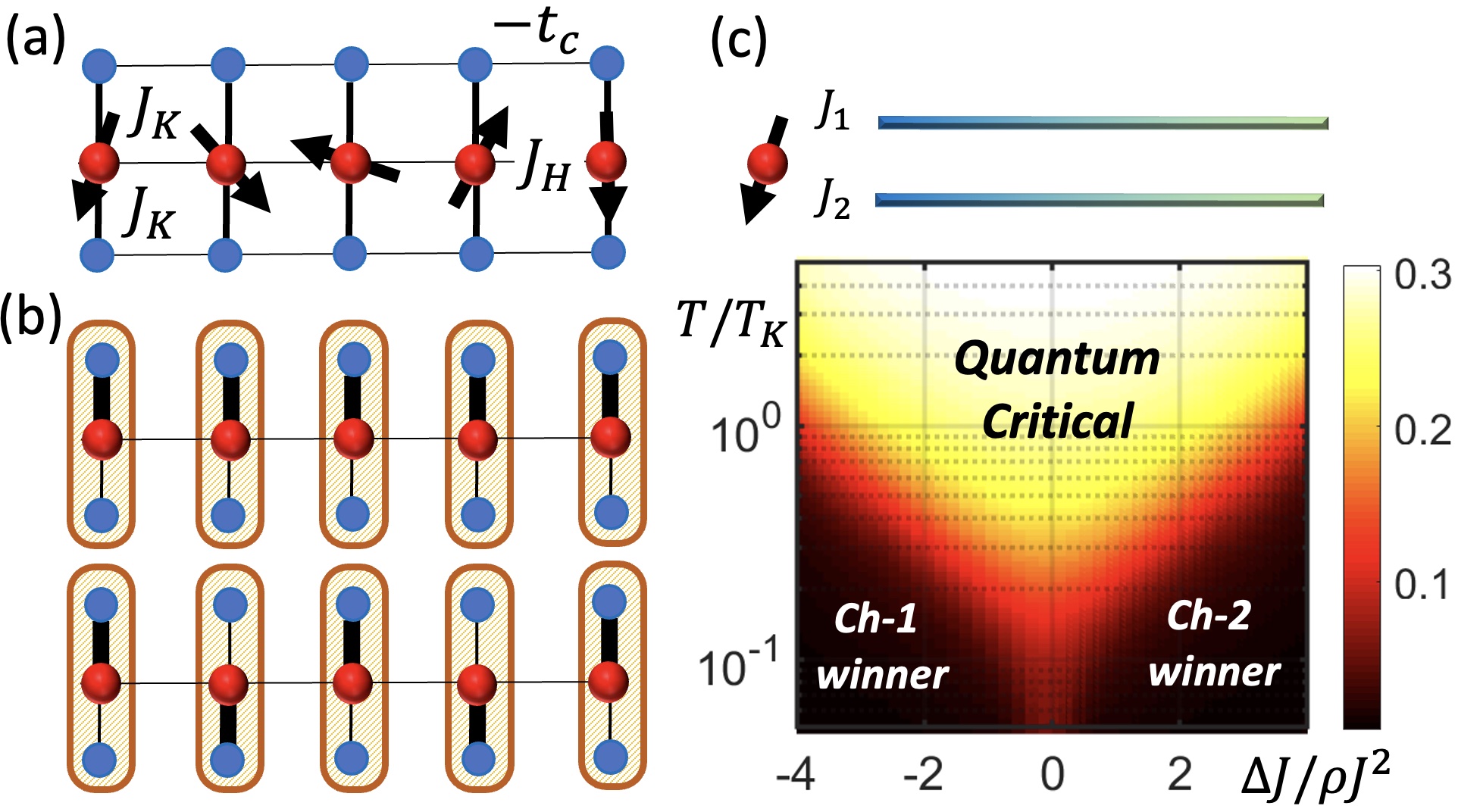}
	\caption{(a) The 1D version of the two-channel Kondo lattice model studied here. (b) The strong coupling leads to a channel magnet; two different patterns of channel symmetry breaking, channel FM (top) and channel AFM (bottom). Bold lines represent spin-singlets. (c) The entropy $S$ of two-channel Kondo impurity vs channel asymmetry and temperature. At the symmetric point, $S$ reduces to a fraction of the high-$T$ value.}\label{fig1}
\end{figure}

The MCKL model is described by the Hamiltonian
\begin{equation}
H=H_c+J_K\sum_{j}\vec S_{j}\cdot c\dg_{ja}\vec\sigma c\dn_{ja} 
\label{eqH}
\vspace{-.2cm}
\end{equation}
where $H_c=-t_c\sum_{\braket{ij}}(c\dg_{i\alpha a}c\dn_{j\alpha a}+\mathrm{H.c.})$ is the Hamiltonian of the conduction electrons and Einstein summation over spin $\alpha,\beta=1\dots N$ and channel $a,b=1\dots K$ indices is assumed. This model has SU($N$) spin and SU($K$) channel symmetries and we are interested in analyzing the effect of a channel symmetry breaking $H\to H+\sum_j\Delta \vec J_j\cdot \vec {\cal O}_j$, where $\vec {\cal O}_j\equiv (\vec S_{j}\cdot c\dg_{ja}\vec\sigma c\dn_{jb}) \vec \tau_{ab}$ and $\vec\tau$'s act as Pauli matrices in the channel space {\cite{Hoshino11}}.
At first look, at least certain deformation of the MCKL can be thought of as a channel magnet. (A na\"ive strong coupling limit is not a spin-singlet, but the {N}ozi\`eres doublet. See supplementary materials \cite{SM} for a deformation that changes this.) In the $J_K\to\infty$ limit \cite{SM}, the spin is quenched due to formation of Kondo singlet with either (for $K=2$) of the channels, leading to a doublet over which $\vec{\cal O}$ acts like $ \vec\tau$~{\cite{Schauerte05,SM}}. Interaction among adjacent doublets leads to a ``channel magnet'' $H_{\rm eff}\propto\frac{t^2}{J_K}\sum_{\braket{ij}}\vec {\cal O}_i\cdot\vec{\cal O}_{j}$. While channel Weiss-field favors a channel anti-ferromagnetic (channel AFM) super-exchange interaction, the mean-field theory predicts a variety of channel ferromagnetic (channel FM) and channel AFM solutions [Fig.\,\ref{fig1}(b)] depending on the conduction filling. 

On the other hand, some differences to a channel magnet are expected since the winning channel has a larger FS \cite{ofc,Wugalter20} and the order parameter $\vec{\cal O}$ is strongly dissipated by coupling to fermionic degrees of freedom. Although a channel-symmetry broken ground state is predicted by both single-site dynamical mean-field theory (DMFT) \cite{Hoshino11,Kuramoto} and static mean-field theory~\cite{vanDyke19,Zhang18,Wugalter20}, it has {not been observed} in recent cluster DMFT studies~\cite{Inui20}. Furthermore, the effective theory of fluctuations in the large-$N$ limit \cite{Wugalter20} 
predicts a disordered phase below the lower critical dimension but the nature of this quantum paramagnet is unclear. In 1D, Andrei and Orignac have used non-Abelian bosonization to show \cite{Andrei2000} that the ground state is gapless and fractionalized (dispersing Majoranas for $K\!=\!2$), {a prediction that contradicts the analysis by Emery and Kivelson \cite{Emery1993}}, and has \emph{not} been confirmed by the density matrix renormalization group calculations \cite{Schauerte05}. 

Resolving these issues requires a technique that is applicable to arbitrary dimensions and goes beyond static mean field and DMFT by capturing both quantum and spatial fluctuations. Here, we show that the dynamical large-$N$ approach, 
recently applied successfully to study Kondo lattices \cite{Rech2006,Komijani18,komijani2019emergent,wang2020quantum,shen2020strange,Wang20,DrouinTouchette2021,Han21}, is precisely such a technique. 

We assume the spins transform as a spin-$S$ representation of SU($N$). 
In the impurity case \cite{Ftnote2}, the spin is fully screened for $K=2S$ whereas it is overscreened and underscreened for $K>2S$ and $K<2S$, respectively \cite{ZinnJustin1998}. The focus of this Letter is on {the Kondo-dominated} regime of the double-screened case $K/2S=2$ which is schematically shown in Fig.\,\ref{fig1}(a). We use Schwinger bosons $S_{j\alpha\beta}=b\dg_{j\alpha}b\dn_{j\beta}$ to form a symmetric representation of spins with the size $2S=b\dg_{j\alpha}b\dn_{j\alpha}$. We then rescale $J_K\to J_K/N$ and treat the model \eqref{eqH} in the large-$N$ limit, by sending $N,K,S\to\infty$, but keeping $s=S/N$ and $\gamma=K/N=4s$ constant. The constraint is imposed on average via a uniform Lagrange multiplier $\mu_b$. 

\begin{figure}[tp!]
	\includegraphics[width=0.99\linewidth]{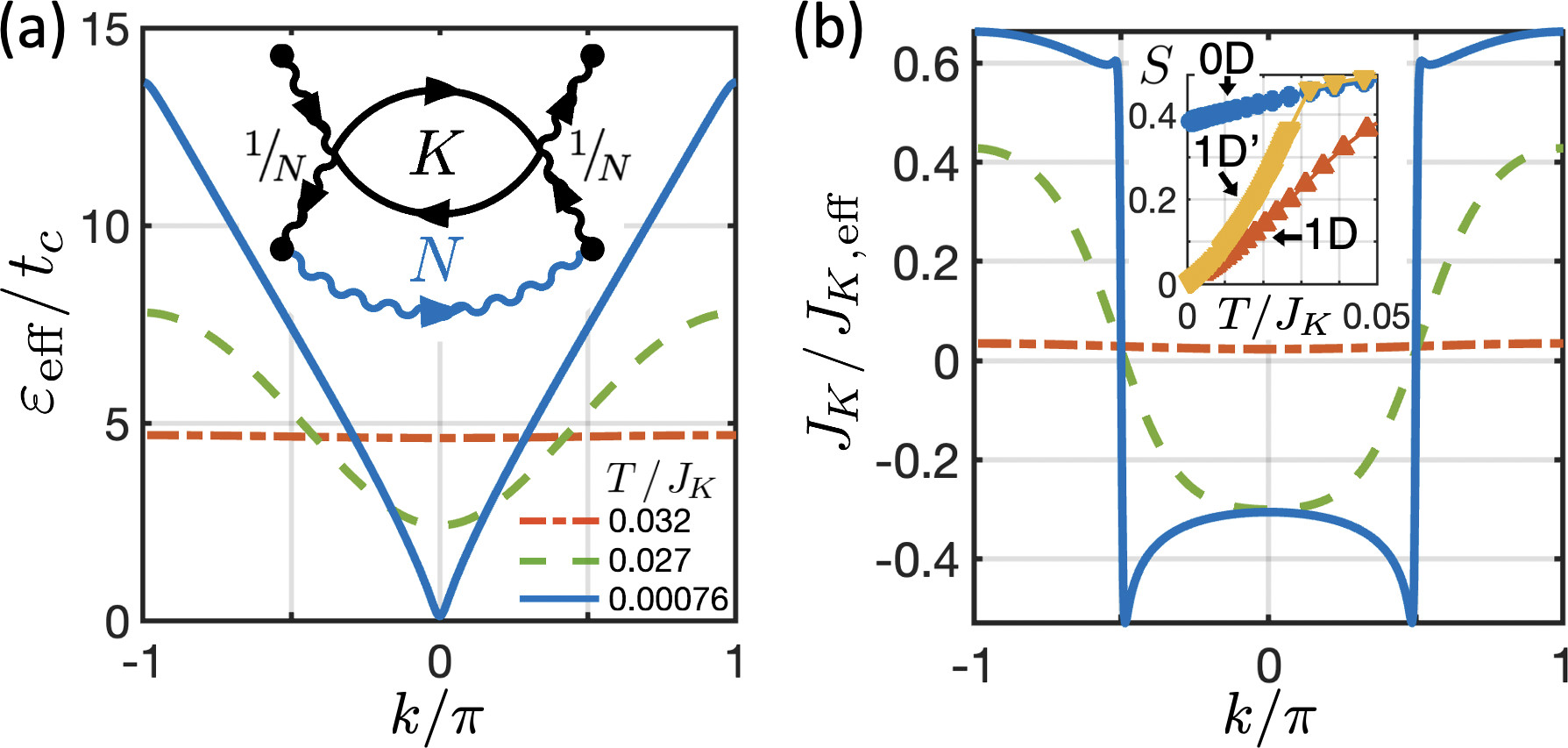}
	\caption{1D 2CKL model. The temperature evolution of (a) the effective energy $\varepsilon_\mathrm{eff}$ for spinons and (b) the inverse effective Kondo coupling $J_{K,\mathrm{eff}}^{-1}$ for holons. At high-$T$, $J_{K,\mathrm{eff}}=J_K$ with no $k$ dependence. Initially, Kondo effect develops locally and $J_{K,\mathrm{eff}}^{-1}\to0$. Then dispersion emerges in both $G_\chi$ and $G_b$, with $J_{K,\mathrm{eff}}^{-1}$ vanishing only at $k\!\sim\!\pm k_F$ and $\varepsilon_\mathrm{eff}$ only at $k\!\sim\!0$. Inset of (a): Despite an $\mathrm{O}(1/N)$ RKKY interaction (black), an initial spinon dispersion (blue) can lead to an O(1) amplification to in the present overscreened case. Inset of (b): Entropy $S$ vs $T$ for 0D, 1D $(t_b=0.2t_c)$, and 1D$'$ $(t_b=0.0002t_c)$. }\label{fig2}
\end{figure}

In the present large-$N$ limit, the Ruderman-Kittel-Kasuya-Yosida (RKKY) interaction is $\mathrm{O}(1/N)$ [inset of Fig.\,\ref{fig2}(a)] and we need to include an explicit Heisenberg interaction $H\to H+J_H\sum_{\braket{ij}}\vec S_i\cdot\vec S_j$ between nearest neighbors $\braket{ij}$ to couple the impurities. Nevertheless, we will show that an infinitesimal $J_H$ is sufficient to produce significant magnetic correlations due to a novel variant of RKKY interaction. For simplicity we limit ourselves to ferromagnetic correlations $J_H<0$. 

For a ${\cal V}$ site lattice, the Lagrangian becomes~\cite{Parcollet1997,Komijani18}
\begin{eqnarray}
{\cal L}&=&\sum_{k}\bar c_{ka\alpha}(\partial_\tau+\eps_k)c_{ka\alpha} +\sum_{k}\bar b_{k\alpha}(\partial_\tau+\varepsilon_k)b_{k\alpha}\label{eqLag}\\
&&\hspace{-.5cm}+\sum_j\frac{\bar \chi_{ja}\chi_{ja}}{J_K}+\sum_{j}\frac{1}{\sqrt N}(\bar \chi_{ja}b_{j\alpha}\bar c_{ja\alpha}+\mathrm{H.c.})+2{\cal V}\mu_b {S}.\nonumber
\end{eqnarray}
Here, $b$'s are bosonic spinons and $\chi$'s are Grassmannian holons that mediate the local Kondo interaction. In momentum space, the electrons and bosons have dispersions $\eps_k=-2t_c\cos k-\mu_c$ and $\varepsilon_k=-2t_b\cos k-\mu_b$, respectively. $t_b$ is the (assumed to be homogeneous) nearest neighbor hopping of spinons due to large-$N$ decoupling of $J_H$ term~{\cite{Komijani18}}. Here, we focus on a half-filled conduction band $\mu_c=0$, but similar results are obtained at other commensurate fillings \cite{SM}. In the large-$N$ limit the dynamics is dominated by the non-crossing Feynman diagrams, resulting in boson and holon self-energies  [$\vec r\equiv (j,\tau)$]
\begin{equation}
\Sigma_b(\vec r)=-\gamma G_c(\vec r)G_\chi(\vec r), \quad \Sigma_\chi(\vec r)=G_c(-\vec r)G_b(\vec r), \label{eqself}
\end{equation}
whereas $\Sigma_c$ is $\mathrm{O}(1/N)$ and thus the electrons propagator $G_c^{-1}(k,{\rm z})={\rm z}-\eps_k$ remains bare, with ${\rm z}$ complex frequency. 
Equations \eqref{eqself} together with the Dyson equations $G_b^{-1}(k,{\rm z})={\rm z}-\varepsilon_k-\Sigma_b(k,{\rm z})$ and $G_{\chi,a}^{-1}(k,{\rm z})=-J_{K,a}^{-1}-\Sigma_\chi(k,{\rm z})$ form a set of coupled integral equations that are solved iteratively and self-consistently, while $\mu_b$ is adjusted to satisfy the constraint. Thermodynamic variables are then computed from Green's functions \cite{Rech2006,Komijani18}.

First, we study the case in which $J_H$ is absent, or $\varepsilon_k=-\mu_b$. In this limit, the self-energies remain local $\Sigma_{b,\chi}(n,\tau)\to\delta_{n0}\Sigma_{b,\chi}(\tau)$ and the problem reduces to the impurity problem \cite{Parcollet1997}. It has never been studied whether the large-$N$ overscreened impurities are susceptible to symmetry breaking \cite{Affleck92}. To do so, we assume that half of $K$ channels are coupled to the impurity with $J_K+\Delta J$ and the other half with $J_K-\Delta J$. This corresponds to a uniform symmetry breaking deformation $\Delta{\cal L}=(\Delta J/J_K^2)\sum_j[\bar\chi_{j1}\chi_{j1}-\bar\chi_{j2}\chi_{j2}]$ of the Lagrangian.
Figure \ref{fig1}(c) shows the entropy of the 2CK impurity model as a function of channel asymmetry, verifying that the impurity is indeed critical with respect to channel symmetry breaking. {In symmetric 2CK, the ground state entropy at large-$N$ is fractional with a universal dependence on $(\gamma,s)$ \cite{Parcollet1997,SM}.}

Next, we focus on finite $t_b$ case for two settings of 1D and $\infty\text{D}$, which correspond to a Bethe lattice with coordination numbers $z=2$ and $z=\infty$. In 1D, $G(k,{\rm z})$ and $\Sigma(k,{\rm z})$ depend on $k$ and ${\rm z}$, but in $\infty\text{D}$, 
self-energies have no spatial dependence and the Green's functions of spinons/electrons obey $G_{b,c}^{-1}={\rm z}+\mu\dn_{b,c}-\Sigma\dn_{b,c}({\rm z})-t_{b,c}^2G\dn_{b,c}$.

\begin{figure}[tp!]
	\includegraphics[width=1\linewidth]{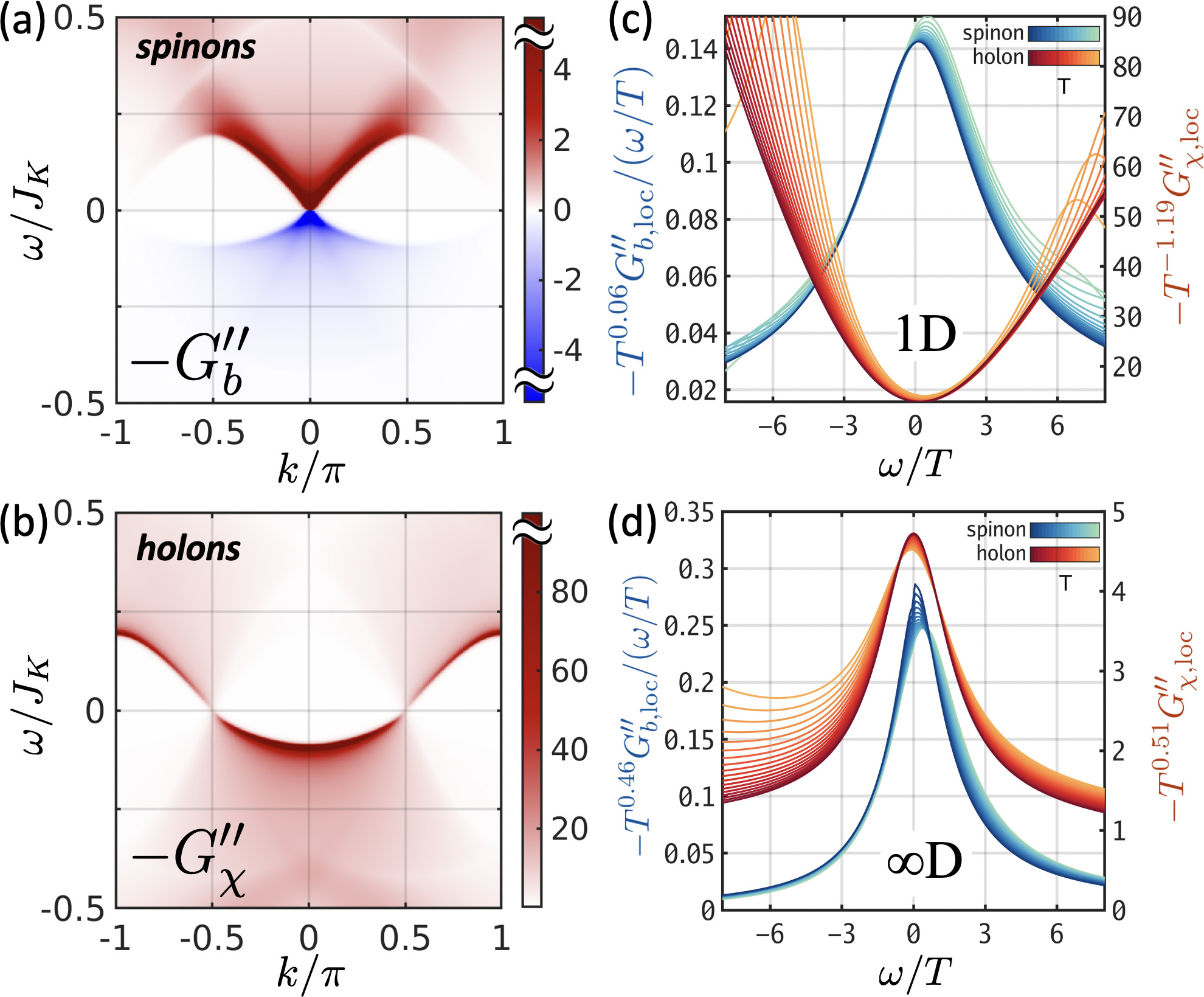}
	\caption{ The spectral function of (a) spinons and (b) holons in a 1D two-channel Kondo lattice at $T/J_K=0.0072$, showing emergent linearly-dispersing spinons at $k=0$ (bare dispersion is quadratic) and holons with Fermi point at $\pm k_F$. Scaling collapse of spinon and holon Green's functions in the 2CK critical regime in (c) 1D lattice ($z=2$) $0.0072 \le T/J_K \le 0.03$ and (d) $\infty$D Bethe lattice ($z=\infty$)  $0.006 \le T/J_K \le 0.03$. For both cases, $J_K/{t_c}=6$, $t_b/{t_c}=0.2$, and $s=0.15$. }\label{fig3}
\end{figure}

Importantly, the criticality of overscreened impurity solution ensures that an infinitesimal spinon hopping seed $t_b\sim 0$ can get an O(1) amplification [inset of Fig.\,\ref{fig2}(a)] and dispersions for spinons and holons are \emph{dynamically} generated. Restricting ourselves to translationally invariant solutions with lattice periodicity ${\rm a}$, this effect can be succinctly represented by the zero-frequency spinon and holon effective dispersion $J_{K,{\rm eff}}^{-1}(k)\equiv -{\rm Re}[G_\chi^{-1}(k,\omega=0)]$ and $\varepsilon_{\rm eff}(k)\equiv-{\rm Re}[G_b^{-1}(k,\omega=0)]$, shown in Figs. \ref{fig2}(a) and \ref{fig2}(b) for various temperatures. This emergent spinon dispersion is independent of the choice of the seed and agree{s} qualitatively with the finite $t_b$ results~\cite{SM}. The consumption of the residual entropy in the lattice by the emerging dispersion is visible in the inset of Fig.\,\ref{fig2}(b). {We stress that in 1D, this apparent transition most likely becomes a crossover when N is finite \cite{Read1985}. In the case of $\infty$D, the system is prone to spin or channel magnetization, as discussed later. Such symmetry breakings would consume the residual entropy \cite{SM}.}

Figures \ref{fig3}(a) and \ref{fig3}(b) shows the finite frequency spectral function of spinons and holons, respectively. Both are dominated by a sharp mode with {emergent} Lorentz invariance. The spinons are gapless and linearly dispersing and the holons form a FS. The temperature collapse of Fig.\,\ref{fig3}(c) confirms that the spectra are critical with the local spectra obeying a $T^{1-2\Delta_{b,\chi}}G''_{b,\chi}(x=0,\omega)=  f_{b,\chi}(\omega/T)$ behavior. Figure \ref{fig3}(d) shows similar collapse for the case of infinite-coordination Bethe lattice ($\infty$D). A marked difference between the two cases is that $\Delta_\chi>1/2$ for 1D, {which leads to $-G''_\chi$ minima at $\omega\sim 0$, whereas $\Delta_\chi<1/2$ in $\infty$D, manifested as a peak at $\omega\sim 0$}. 

\begin{figure}
\includegraphics[width=1\linewidth]{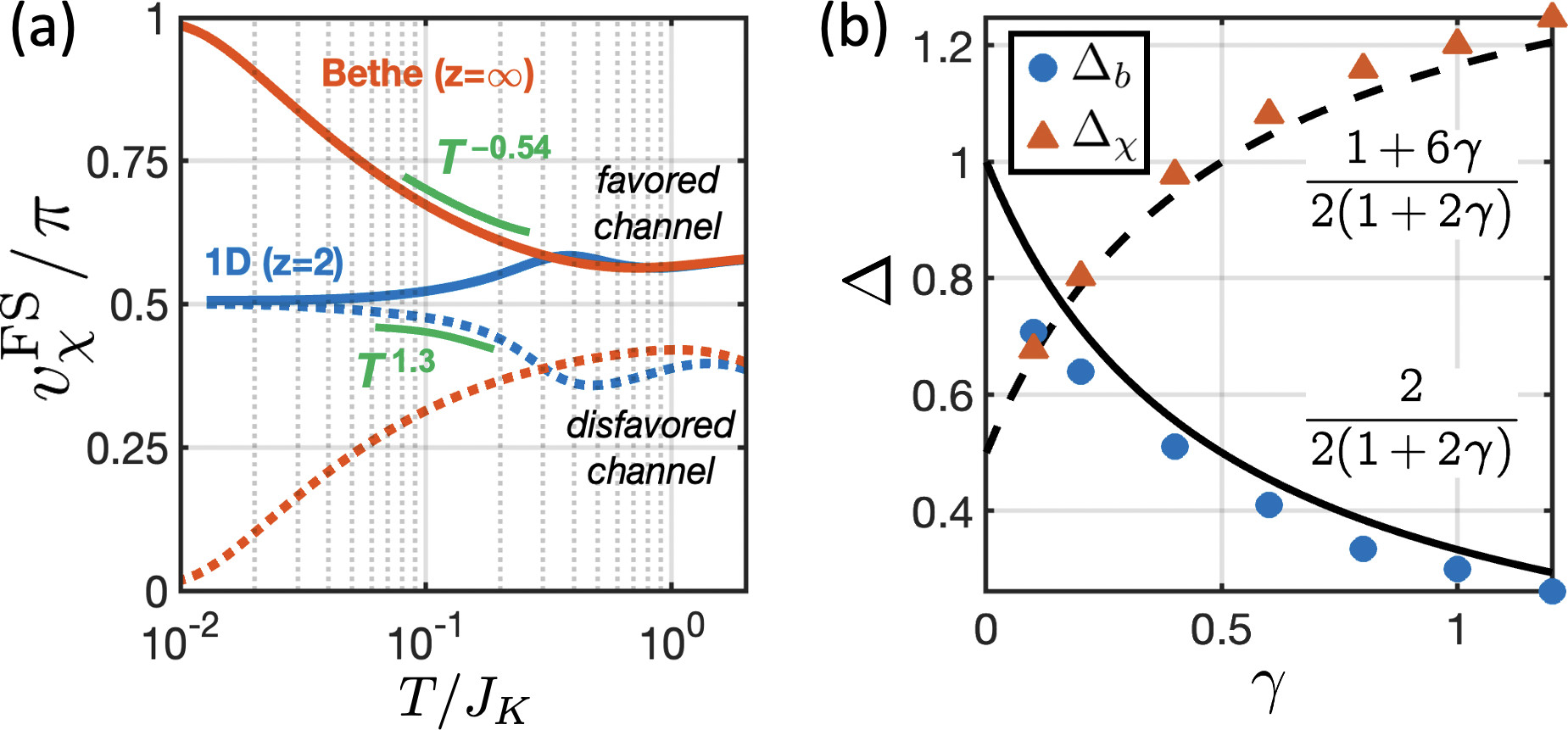}
\caption{
	(a) The evolution of the FS in the presence of small channel symmetry breaking in 1D and $\infty$D with temperature. (b) The scaling exponents $\Delta_{b/\chi}$ in 1D from the numerics. The lines show the analytical values given by Eqs.\,\eqref{expo1D}.
}\label{fig4}
\end{figure}

{What is the effect of channel symmetry breaking on the volume of FS?} According to Luttinger's theorem, the FS volume is related to electron phase shift $v_{a}^{\mathrm{FS}}={\cal V}^{-1}\sum_k{\delta_{a}(k)}$ {for a $d$ dimensional lattice}. From $K=4S$ case of the Ward identity \cite{Coleman2005}, the electron phase shift is related to that of holons $N\delta_{c,a}(k)=\delta_{\chi,a}(k)$, which itself is defined as
\begin{equation}
\delta_{\chi,a}(k)=-{\rm Im}\{\log[-G_{\chi,a}^{-1}(k,0+i\eta)]\}.\label{eqdelta}
\end{equation}
The locus of points at which $J^{-1}_{K,{\rm eff}}(k)$ changes sign defines a holon FS which generalizes to any dimension. In 1D, holons are occupied for $\abs{k}<\pi/2$. So,
we find that $v_{\chi,a}^{\mathrm{FS}}=2\pi S/K=\pi/2$ and the total change in electron FS is {$N\Delta v_{c,a}^{\mathrm{FS}}=\pi/2$}, corresponding to a large FS in the critical phase.
We use Eq.\,\pref{eqdelta} to study the effect of a uniform symmetry breaking field $\Delta{\cal L}$. Figure \ref{fig4}(a) shows how FSs of slightly favored and disfavored channels evolve as a function of $T$ in the two cases. In 1D, the FS asymmetry disappears, restoring a channel symmetric criticality at low $T$, {consistent with the Mermin-Wagner theorem}. On the other hand, in $\infty$D the asymmetry grows and one channel totally decouples from the spins, with gapped spinons and also gapped holons for both channels. The exponents are related to $\Delta_\chi$; varying $\Delta J$ in Eq.\,\pref{eqdelta} we find
{
\be
\frac{\partial {v_{\chi,a}^{\mathrm{FS}}}}{\partial {\Delta J}}=\frac{-1}{\cal V}\sum_kG''_{\chi}(k,0+i\eta) =-G''_{\chi}(x=0,0+i\eta).
\ee
Assuming $\abs{G_\chi (\vec r)}\sim \abs{\vec r}^{-2\Delta_\chi}$, the holon FS is unstable against symmetry breaking when $G''_\chi(k_F,0+i\eta)\sim T^{2\Delta_\chi-d-1}$ diverges. This $2\Delta_\chi<d+1$ regime coincides with when the symmetry breaking term $\Delta J$ is relevant, in the renormalization group (RG) sense. On the other hand instability of the entire holon FS requires the divergence of $G''_\chi(x=0,0+i\eta)\sim T^{2\Delta_\chi-1}$, i.e. $2\Delta_\chi<1$ which is a more stringent condition and agrees} with Fig.\,\ref{fig4}(a), confirming $\Delta_\chi=1/2$ as the marginal dimension. 

Figure \ref{fig4}(a) shows that the symmetry breaking $\Delta{\cal L}$ is relevant in $\infty$D, but is irrelevant in 1D. To establish this from the microscopic {model}, one has to access the infrared (IR) fixed point. From the numerics we see that the system flows to a critical IR fixed point, in which spinons and holons are critical in addition to electrons. For an impurity $G_b\sim \abs{\tau}^{-2\Delta_b}$ and $G_\chi(\tau)\sim \abs{\tau}^{-2\Delta_\chi}$ are reasonable at $T=0$. The exponents are known \cite{Parcollet1997,SM}:
\begin{equation}
0,\infty{\rm D}:\qquad \Delta_\chi=\frac{\gamma}{2(1+\gamma)},\qquad \Delta_b=\frac{1}{2(1+\gamma)},\label{eq0D}
\end{equation}
and coincide with those of the $\infty$D in the small $t_b$ regime we are interested here \cite{SM}. 
In the presence of a dimensionless $\lambda_0=\Delta J/\rho J_K^2$, the
RG analysis $d\lambda/d\ell=(1-2\Delta_\chi)\lambda$ predicts a dynamical scale $w\sim T_K\lambda_0^{1+\gamma}$ [cf. Fig.\,\ref{fig1}(c)].

The 1D case is more subtle; as $T\to 0$, we see from Fig.\,\ref{fig2} that $J^{-1}_{K,\mathrm{eff}}(\pm k_F)\to 0$ and $\varepsilon_{\rm eff}(0)\to 0$ at the IR fixed point \cite{Ftnote3}. This means that the Kondo coupling flows to strong coupling at $\abs{k}<k_F$, to weak coupling at $\abs{k}>k_F$, and gets critical at $k=\pm k_F$, while the spinons are gapless at $k\!=\!0$.
At these momenta, the Dyson equation has the scale-invariant form $G_b\Delta\Sigma_b\vert_{k\sim 0}=G_\chi\Delta\Sigma_\chi\vert_{k\sim\pm k_F}=-1$.

We can obtain a low-energy description by expanding fields near zero energy, e.g. 
$\psi(x)\sim e^{ik_Fx}\psi_R+e^{-ik_Fx}\psi_L$ for electrons and holons. In 1+1 dimensions, the conformal invariance of the fixed point dictates the following form for the $T=0$ Green's functions $G(x,\tau)=G(z,\bar z)$:
\begin{equation}
G_b=-\bar\rho\Big(\frac{\rm a^2}{\bar zz}\Big)^{\Delta_b}\hspace{-.25cm}, \quad G_{\chi R/L}=\frac{-1}{2\pi}\Big(\frac{\rm a}{\bar z}\Big)^{\Delta_\chi\pm \frac12}\Big(\frac{\rm a}{z}\Big)^{\Delta_\chi\mp \frac12}\hspace{-.15cm}\label{eqG}
\end{equation}
where $z=v\tau+ix$ and $\bar \rho=2s/{\rm a}$. The $G_{cR/L}$ is obtained from $G_{\chi R/L}$ by $\Delta_\chi\to 1/2$. These Green's functions can be conformally mapped to finite-$T$ via $z\to ({\beta}/{\pi})\sin({\pi z}/{\beta})$ replacement. Furthermore, in terms of $q=k+i\omega/v$, they have the Fourier transforms:
\begin{eqnarray}
G_b&=&-2\pi {\rm a}^2 \bar\rho v_b^{-1}({\rm a}^2\bar qq)^{\Delta_b-1}\zeta_0(\Delta_b)\label{eqGq}\\
G_{\chi R/L}&=&\mp {\rm a}^2v_\chi^{-1}({\rm a}\bar q)^{\Delta_\chi-1\mp 1/2}({\rm a} q)^{\Delta_\chi-1\pm1/2}\zeta_1(\Delta_\chi)\nonumber
\end{eqnarray}
where $\zeta_n(\Delta)\equiv2^{1-2\Delta}\Gamma(1-\Delta+n/2)/\Gamma(n/2+\Delta)$.
From matching the powers of frequency in Eqs.\,\eqref{eqself}, \eqref{eqG} and \eqref{eqGq}, we conclude that $\Delta_b+\Delta_\chi=3/2$ in order to satisfy the self-consistency. Moreover, from the matching of the amplitudes of the Green's functions we find~\cite{SM}
\begin{equation}
{\rm 1D}:\qquad\Delta_\chi=\frac{1+6\gamma}{2(1+2\gamma)}, \qquad \Delta_b=\frac{2}{2(1+2\gamma)}.\label{expo1D}
\end{equation}
Note that $\Delta_\chi>1/2$, ensuring that channel symmetry breaking perturbations are irrelevant in 1D. These are in excellent agreement with the exponents extracted from $\omega/T$ scaling [Fig.\,\ref{fig4}(b)] and we have established a semianalytical framework to interpolate between 1D and $\infty\text{D}$. 

The emergent dispersion in Fig.\,\ref{fig2}, the scaling dimensions in Eq.\,\eqref{expo1D}, and their relation to symmetry breaking in Fig.\,\ref{fig4} are the central results of this Letter. In the following we discuss some {of the} implications of these results {for physical observables that are independent of our fractionalized description, leaving the details to \cite{SM}.}

The fractionalization $S_{\alpha\beta}\sim b\dg_\alpha b\dn_\beta$ or $b\dg_\alpha c\dn_{a\alpha}\sim \chi_a$ contraction are related to order parameter fractionalization~\cite{ofc,ofc2}. In the long time/distance limit, correlation functions of $b\dg_\alpha c\dn_{a\alpha}$ and that of $\chi\dn_a$ are given by $\Sigma_\chi$ and $G_\chi$, respectively and thus, have exponents that add up to zero. On the other hand, correlators of gauge-invariant  operators {${\cal X}_{ab}\equiv \bar\chi_a\chi_b$} and ${\cal O}_{ab}\equiv b\dg_\alpha b\dn_{\beta} c\dg_{b\beta}c\dn_{a\alpha}$ are exactly equal since both can be constructed by taking derivatives of free energy with respect to $
\Delta J^{ab}$, either before or after Hubbard-Stratonovitch transformation. A diagrammatic proof of this equivalence is provided in \cite{SM}. Scaling analysis gives $\chi_{\mathrm{ch}}(x=0)\sim T^{4\Delta_\chi-1}$ and $\chi_{\mathrm{ch}}^{\rm 1D}(q=0)\sim T^{4\Delta_\chi-2}$ up to a constant shift coming from the regular part of free energy. 

Another nontrivial feature of 2CK impurity fixed point is its magnetic instability \cite{Affleck92} whose large-$N$ incarnation is $\Delta_b<1/2$ for the impurity (or $\infty\text{D}$) in Eq.\,\eqref{eq0D}. From Eq.\,\eqref{expo1D}, we see that this also holds for 1D 2CKL for $\gamma>1/2$. This is reflected in the divergence  of the uniform $\chi_m(q=0)$ static magnetic susceptibilities as a function of $T$. 
Using scaling analysis $\chi_m^{\mathrm{1D}}(q=0)\sim T^{4\Delta_b-2}$ and $\chi_m(x=0)\sim T^{4\Delta_b-1}$ up to a constant shift, in good agreement with numerics \cite{SM}. {Note that this critical spin behavior is different from the gapped spin sector observed in \cite{Emery1993,Schauerte05}, but is qualitatively consistent with \cite{Andrei2000}.}

Lastly, the fact that the fixed point discussed above is IR stable follows from the fact that the interaction is exactly marginal due to $\Delta_b+\Delta_\chi=3/2$ and that vertex corrections remain $\mathrm{O}(1/N)$. The 1+1D correlators \pref{eqG} can be obtained from three sets of decoupled Luttinger liquids for each of the $c,b,\chi$ fields with fine-tuned Luttinger parameters that give the correct exponents. Such a spinon-holon theory will have a Virasoro central charge $c_0/N=1+\gamma$. On the other hand the coset theory of \cite{Andrei2000,Azaria1998,SM} predicts $c_{AO}/N=\gamma/(1+\gamma)$. We have used $T\to 0$ heat-capacity and the excitation velocities $v$ to compute the central charge according to $C/T=(\pi k_B^2/6v)c$ as a function of $\gamma$ and found $c=c_0$ \cite{SM}. Note that there is no contradiction with the $c$-theorem since the UV theory is not Lorentz invariant due to ferromagnetism. The discrepancy with $c_{AO}$ is likely rooted in inability of Schwinger bosons to capture gapless spin liquids \cite{Arovas1988}.

In summary, we have shown that the dynamical large-$N$ approach can capture symmetry breaking in multichannel Kondo impurities and lattices in the presence of both emergent and induced ferromagnetic correlations within an RG framework with explicit examples on 0D, 1D, and $\infty$D. The scaling analysis enables an analytical solution to the critical exponents and susceptibilities which are in good quantitative agreement with numerics, and is applicable to higher dimensional CFTs. A determination of the upper and lower critical dimensions and the effect of antiferromagnetic correlations are left to a future work \cite{Yang2}. 

\emph{Acknowledgment}---The authors acknowledge fruitful discussions with P.~Coleman and N.~Andrei. This work was performed in part at Aspen Center for Physics, which is supported by NSF Grant No.~PHY-1607611. Computations for this research were performed on the Advanced Research Computing Cluster at the University of Cincinnati, and the Penn State University's Institute for Computational and Data Sciences' Roar supercomputer.

\bibliography{KL}

\begin{thebibliography}{46}
\bibitem [{\citenamefont {{We use 0{D} to refer to the impurity
  problem}}()}]{Ftnote2}%
  \BibitemOpen
  \bibfield  {author} {\bibinfo {author} {\bibnamefont {{We use 0{D} to refer
  to the impurity problem}}},\ }\href@noop {} {\bibinfo  {journal} {although a
  conduction band exists}\ }\BibitemShut {NoStop}%
\bibitem [{\citenamefont {{This is not the case for the spinon's $\pi$-mode,
  $\varepsilon_{\rm eff}(\pm\pi/{\rm a})\neq 0$}}()}]{Ftnote3}%
  \BibitemOpen
  \bibfield  {author} {\bibinfo {author} {\bibnamefont {{This is not the case
  for the spinon's $\pi$-mode, $\varepsilon_{\rm eff}(\pm\pi/{\rm a})\neq
  0$}}},\ }\href@noop {} {\bibinfo  {journal} {but this mode has negligible
  spectral weight}\ }\BibitemShut {NoStop}%
\end{thebibliography}

\appendix

\section{Supplementary materials}

\subsection{1. Strong coupling and channel magnet}
We can use SU$_{\mathrm{sp}}$(2)$\otimes$SU$_{\mathrm{ch}}$(2)$\otimes$U(1) symmetries to label various local states of the two-channel Kondo lattice. First, we consider a single-site two-channel Kondo model. The Hamiltonian
\begin{equation}
	H_{\mathrm{2CK}}/J_K=\vec S\cdot (c\dg_1\vec \sigma c\dn_1+c\dg_2\vec \sigma c_2)
\end{equation}
can be written as
\begin{equation}
	H_{\mathrm{2CK}}/J_K=-\frac{11}{4}+\vec S^2_{\mathrm{tot}}+\vec C^2+\frac{1}{2}Q^2.
\end{equation}
in terms of the charge and Casimirs of spin and channel 
\begin{equation}
	\vec S_{\mathrm{tot}}=\vec S+c\dg\frac{\vec\sigma}{2}c, \quad \vec C=c\dg\frac{\vec\tau}{2}c,\quad Q=c\dg c-2,
\end{equation}
which are $\vec S^2=S(S+1)$ and $\vec C^2=C(C+1)$. The energies of the 32 resulting states are listed in Table~\ref{table}. The ground state is the Nozi\`eres doublet corresponding to the overscreened state. This Hamiltonian can be deformed to
\begin{equation}
	H_{\mathrm{2CK}}^{\rm deformed}/J_K=H_{\mathrm{2CK}}/J_K-r_1\vec C^2-r_2Q,
\end{equation}
while preserving the symmetries of the Hamiltonian. $r_1>2/3$ is sufficient to change the ground state to the four spin-singlet channel-doublet states. These are states in which the impurity spin forms a spin-singlet with one of the channels. The remaining channel can be either empty or fully occupied, giving rise to the quartet. This quartet can be split by having a non-zero $r_2$. For example $r_2<0$ will select a channel doublet with the other channel empty, and can be represented as 
\begin{equation}
	\ket{\mu=1/2}=\frac{\ket{\Ua\da_1-\Da\ua_1}}{\sqrt 2},\quad \ket{\mu=-1/2}=\frac{\ket{\Ua\da_2-\Da\ua_2}}{\sqrt 2}.\nonumber
\end{equation}
The operators $\hat{\cal O}^\mu=\vec S\cdot c\dg_a\tau^\mu_{ab}\vec\sigma c_b$ act like Pauli matrices in the space of the doublet $\ket{\pm 1/2}$,
\begin{equation}
	-\frac{2}{3}{\cal O}^\mu\sim \tau^\mu, \qquad \mu=x,y,z.
\end{equation}
Having singled-out a channel doublet, we consider a two-channel Kondo lattice where these local ground states are coupled via the electron-hopping term. We can write
\begin{eqnarray*}
	c\dn_{ia\sigma}\ket{\mu}&=&-\delta_{a\mu}\frac{\tilde\sigma}{\sqrt 2}\ket{\Updownarrow_\sigma},\\
	c\dg_{ia\sigma}\ket{\mu}&=&\delta_{a\mu}\frac{\tilde\sigma}{\sqrt 2}\ket{ch. T^{\tilde a},\sigma}\\
	&&\hspace{1cm}+\delta_{a\bar\mu}[\tilde a\frac{\sqrt 3}{2}\ket{OS,\sigma}+\frac{1}{2}\ket{ch. T^0,\sigma}].
\end{eqnarray*}
where $\tilde a\equiv\sgn(a)$ and we have defined the Nozi\'eres overscreened excited states ($E=-2J_K$)
\begin{eqnarray}
	\ket{OS,+1/2}&=&\frac{1}{\sqrt 6}\ket{\Ua(\ua_1\da_2+\da_1\ua_2)-2\Da\ua_1\ua_2}\!,\\
	\ket{OS,-1/2}&=&\frac{1}{\sqrt 6}\ket{2\Ua\da_1\da_2-\Da(\ua_1\da_2+\da_1\ua_2)}\!.
\end{eqnarray}
and channel triplet excited states $\vert{ch. \vec T,\sigma}\rangle$. For the sake of this section, we can project out the latter assuming that their energy is pushed further up. Therefore, the lowest excited states are the Nozi\'eres states and the empty states $\ket{\Updownarrow_\sigma}$. Treating the hopping via 2nd order perturbation theory, 
\begin{equation}
	H_{\rm eff}=-{\cal P}\Big[H_t{\cal Q}\frac{1}{\Delta H_0}{\cal Q}H_t\Big]{\cal P},
\end{equation}
where ${\cal P}$ and ${\cal Q}$ are projectors to ground state and excited states, respectively and ${\cal P}+{\cal Q}=1$. Assuming that the sites $i$ and $j$ are initially at $\ket{\mu_i,\nu_j}$ and eventually at $\ket{\mu'_i,\nu'_j}$ we find after a straightforward calculation
\begin{equation}
	\bra{\mu'_i,\nu'_j}H_{\rm eff}\ket{\mu_i,\nu_j}=\frac{3t^2}{4\Delta E}\bb X_{\mu\nu}^{\mu'\nu'}, \quad \bb X_{\mu\nu}^{\mu'\nu'}=-\tilde\nu\tilde\nu'\delta_{\bar\mu\nu}\delta_{\bar\mu'\nu'}\nonumber
\end{equation}
where $\Delta E=(1+3r_1/2)J_K$. It can be easily shown that $\bb X=\bb P-\bb 1$ where $\bb P_{\mu\nu}^{\mu'\nu'}=\delta_\mu^{\nu'}\delta_\nu^{\mu'}$ is the exchange matrix and we have the relation
\begin{equation}
	2\bb P=\bb 1+\vec\tau_i\cdot\vec\tau_j.
\end{equation}
This completes the derivation.
\begin{table}
	\begin{tabular}{ccccc}
		\hline\hline
		$S$& $C$ &  $Q$  &  \#  &  E \\
		\hline
		1 & 0 & 0 & 3 & 0\\
		\hline
		1/2 & {1/2} & {$\pm$1} & 8 & 0\\
		\hline
		0 & {0} & {$\pm$2} & 2 & 0\\
		0 & {1} & {0} & 3 & 0\\
		\hline\hline
	\end{tabular}
	\qquad
	\begin{tabular}{ccccc}
		\hline\hline
		$S$ & $C$ &  $Q$ & \# & E \\
		\hline
		3/2 & 0 & 0 & 4 & 1\\
		\hline
		1 & {1/2} & {$\pm$1} & 12 & 1/2 \\
		\hline
		1/2 & {0} & {$\pm$2} & 4 & 0\\
		1/2 & {1} & {0} & 6 & 0\\
		\hline
		0 & {1/2} & {$\pm$1} & 4 & -3/2\\
		\hline
		1/2 & 0 & 0 & 2 & -2\\
		\hline\hline
	\end{tabular}
	\caption{The spectrum of a single-site 2CK model using a SU$_{\mathrm{sp}}$(2)$\otimes$ SU$_{\mathrm{ch}}$(2)$\otimes$U(1) symmetry labeling, (left) free electron (right) after coupling to the spin. $S$, $C$ are the total spin and channel of the state and $Q$ is the charge.}\label{table}
\end{table}

\subsection{2. Details of numerical computation}
Here we describe our algorithm to compute the low temperature Green's functions. All other physical quantities can be computed from them thanks to the large-$N$ limit. The Green's functions are computed from the self-consistency equations,
\begin{eqnarray}
	\Sigma_b(j,\tau)&=&-\gamma G_c(j,\tau)G_\chi(j,\tau),\label{app:self-cons-1}\\
	\Sigma_\chi(j,\tau)&=&G_c(-j,-\tau)G_b(j,\tau). \label{app:self-cons-2}
\end{eqnarray}
We found that they are best represented in momentum and frequency domain. The solutions are solved on linear or logarithmic frequency grids and a linear momentum grid. Using that $G_c^{-1}(k,{\rm z})={\rm z}-\eps_k$, we find
\begin{eqnarray}
	\Sigma''_b(k,\omega)\!\!&=&\!\!-\frac{\gamma}{\mathcal{V}}\sum_p[f(\eps_p)-f(\eps_p-\omega)]G''_\chi(k-p,\omega-\eps_p),\nonumber\\
	\Sigma''_\chi(k,\omega)\!\!&=&\!\!\frac{1}{\mathcal{V}}\sum_p[f(\eps_p)+n_B(\eps_p+\omega)]G''_b(k+p,\omega +\eps_p).\nonumber\label{app:self-cons-SE}
\end{eqnarray}
Thus, only a single sum over momentum is needed. Then we used Hilbert transform to recover the full self-energies, up to a constant in $\Sigma_b$ as discussed later this section. Next, the retarded Green's functions are computed from the self-energies with $\mathrm{z}$ set to $\omega+i\eta$ in
\begin{eqnarray}
	G_b(k,\mathrm{z})&=&\frac{1}{\mathrm{z} -\varepsilon_k-\Sigma_b(k,\mathrm{z})},\\
	G_{\chi}(k,\mathrm{z})&=&\frac{1}{-1/J_{K}-\Sigma_\chi(k,\mathrm{z})},\label{app:self-cons-G}
\end{eqnarray}
A search for $\mu_b$ is lastly used to satisfy the constraint
\begin{equation}
	2s = -\frac{1}{\mathcal{V}}\sum_k\int \frac{\mathrm{d}\omega}{\pi} n_B(\omega) G_b^{\prime\prime}(k,\omega+i\eta).
\end{equation}
The procedures above constitute the essential step in updating the self-consistency equations. Our main solver program is organized as follows:
\begin{enumerate}
	\setcounter{enumi}{-1}
	\item Initialize the self-energies to zero at a high temperature $T$. Also initialize $\mu_b$, say, to $T\log(1+1/2s)$.
	\item At present temperature $T$, initialize $\eta$ to a large value, say $T$. Then, \label{prog:solve-one-T}
	\begin{enumerate}
		\item update the self-consistency equations for $\Sigma_\chi$, $G_\chi$, and $\Sigma_b$; \label{prog:update-selfconsis-1}
		\item search for a $\mu_b$ that gives a $G_b$ satisfying the constraint; then \label{prog:update-selfconsis-2}
		\item reduce $\eta$ and repeat (a--b) until $\eta$ is small compared to $T$, say $\eta=T/32$; then
		\item repeat (a--b) until convergence.
	\end{enumerate}
	\item Reduce $T$ and rerun Step \ref*{prog:solve-one-T}. Repeat until the desired temperature is reached.
\end{enumerate}
Decreasing $\eta$ and $T$ slowly helps with the convergence. At low temperatures, frequency and momentum resolutions limits the convergence. The frequency grid need to be fine enough to resolve the sharp features due to small $T$ and $\eta$. For us typically $\eta/\Delta\omega>7$. The momentum resolution, or finite size effect, limits the lowest $T$ attainable to the order of the Fermi velocity of conduction electrons $v_c/L$, $L$ being the linear dimension.

Strictly speaking, $G_\chi(k,\mathrm{z})$ does not obey a Kramers-Kronig relation, whereas $G_\chi-g_\chi$ does. Consequently, $\Sigma_b$ obtained above is missing a real temperature-dependent constant $\gamma J_K \sum_p f(\eps_p)/\mathcal{V}$. This can be conveniently absorbed into $\mu_b$ and need not be computed.

To improve efficiency, 1D calculations can utilize inversion symmetry. Thus, only half of the momentum grid is needed. At low temperatures, the computation can be further sped up with a frequency grid that is uniform at low frequencies and logarithmic at higher ones.

\begin{figure}[ht]
	\includegraphics[width=1\linewidth]{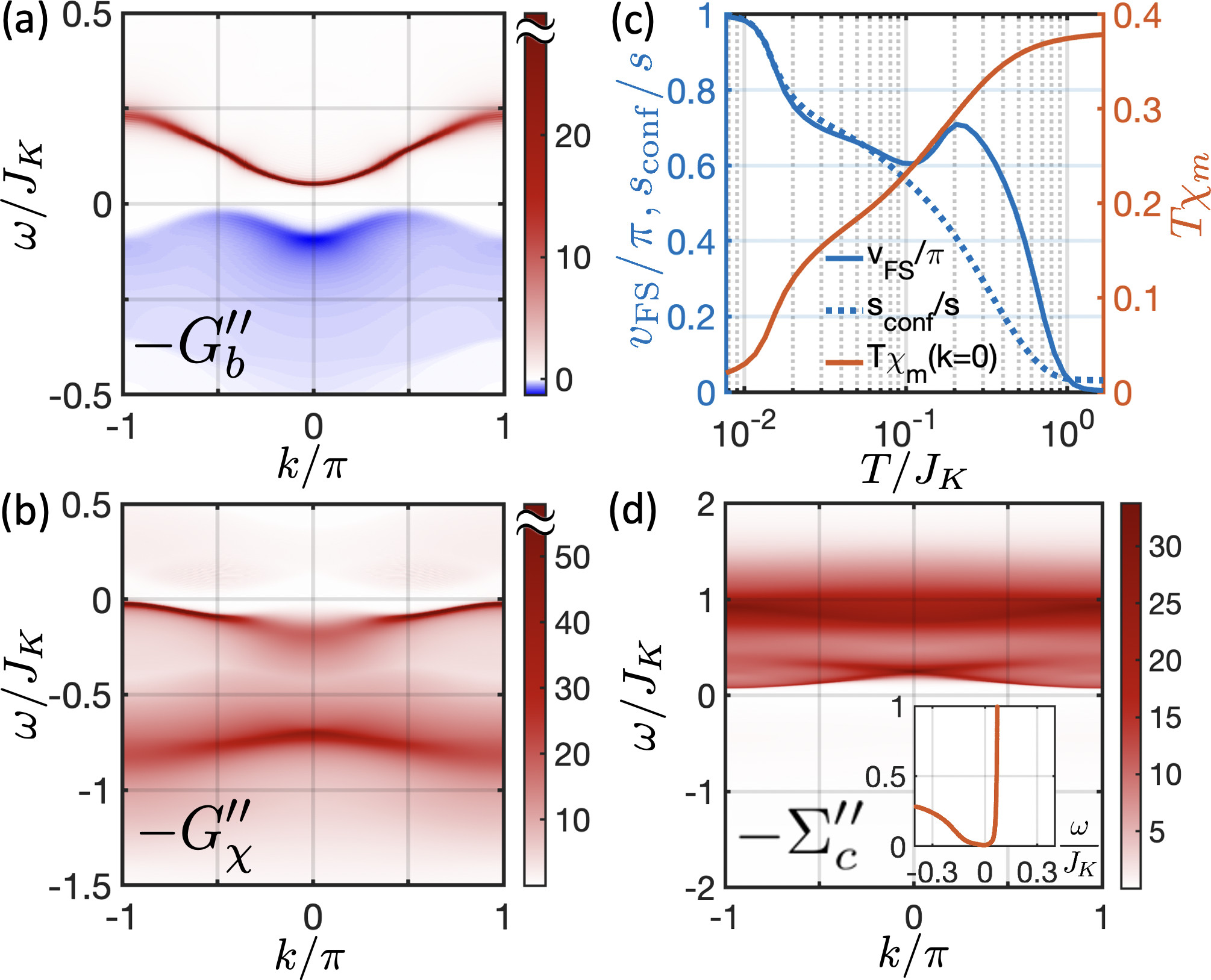}
	\caption{Perfectly screened Kondo lattice with a finite $t_b=0.2 t_c$ and $J_K=6t_c$. The spectral weights of (a) spinons and (b) holons show dispersion on top of a gap. In (c), both the holon Fermi volume, $v_\chi^{\mathrm{FS}}$, and the fraction of spinons at negative frequencies, $s_{\text{conf}}/s$, show the onset of Kondo physics. The uniform magnetic susceptibility is also displayed as $T \chi_m$, showing Curie and Pauli susceptibility at high and low temperatures, respectively. (d) The imaginary part of the self-energy of conduction electrons $-N\Sigma_c^{\prime\prime}$. The inset shows the gap in $-N\Sigma_{c,\mathrm{loc}}^{\prime\prime}$ near zero frequency. Panels (a),(b) and (d) are evaluated at $T=0.0077J_K$.} \label{fig5}
\end{figure}

\subsection{3. {Other screenings and fillings in 1D}}
Here, we briefly present two other screening or filling cases in addition to the one presented in the Letter. 

The first is the fully-screened case $K=2S$ with half-filled conduction electrons,  shown in Fig.\,\ref{fig5}. This has to be contrasted to the Fermi liquid regime of the ferromagnetically coupled Kondo lattice treated in the independent bath approximation in Ref.\,\onlinecite{Komijani18}. The spectrum of spinons $-G_b''(k,\omega+i\eta)$ and holons $-G_\chi''(k,\omega+i\eta)$ [Fig.\,\ref{fig5}(a,b)] show that ground state is gapped. Figure \ref{fig5}(c) shows that the ground state is paramagnetic, all spinons are confined to negative frequencies (as expected from spinon gap) and the electrons have a large FS, as $v^{\mathrm{FS}}_{\chi,a}=\pi$. The plateau in $v^{\mathrm{FS}}_{\chi,a}$ is due to van Hove singularity of the conduction band density of states. Figure \ref{fig5}(d) shows the conduction electron self-energy $-N\Sigma''_c(k,\omega+i\eta)$,
\begin{eqnarray}
	-N\Sigma''_{c,\mathrm{loc}}(\omega+i\eta)&=&\int{\frac{dx}{\pi}[f(x)+n_B(x+\omega)]}\\
	&&\hspace{2.6em}G''_{\chi,\mathrm{loc}}(x)G''_{b,\mathrm{loc}}(x+\omega).\nonumber
\end{eqnarray}
At zero temperature we can simplify this expression to:
\begin{equation}
	-N\Sigma''_{c,\mathrm{loc}}(\omega+i\eta)=\int_{-\omega}^0{\frac{dx}{\pi}}G''_{\chi,\mathrm{loc}}(x)G''_{b,\mathrm{loc}}(x+\omega).
\end{equation}
For the fully-screened case, spinons and holons develop a gap $E_{\mathrm{gap}}\sim T_K$ in their spectrum at zero temperature. Therefore, for $\abs{\omega}\ll E_{\mathrm{gap}}$ this expression is equal to zero and $-\Sigma''_{c,\mathrm{loc}}$ is also gapped. This agrees with the numerical results shown in Fig.\,\ref{fig5}(d).

\begin{figure}
	\includegraphics[width=.95\linewidth]{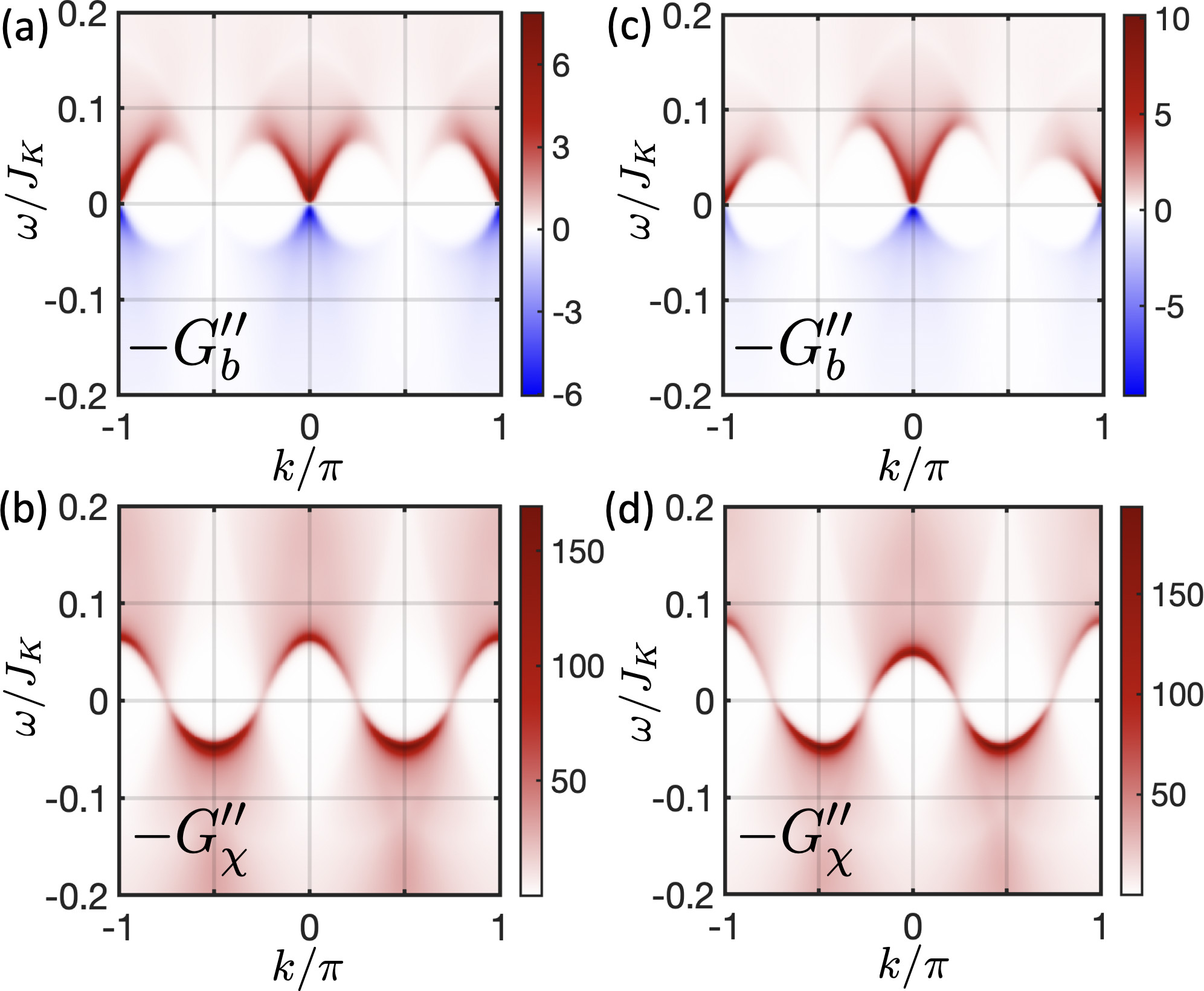}
	\caption{Double-screened Kondo lattice $(K/2S=2)$ with $3/4$-filled conduction electrons and spinon hoppings that are: (a--b) infinitesimal, $t_c/t_b=0.002t_c$; (c--d) finite, $t_b=0.2t_c$.  The spectral weights of spinons and holons show emergent Lorentz invariant modes similar to the half-filled case, but a doubled unit cell under infinitesimal $t_b$. The case of finite $t_b$ breaks the symmetry between modes around $k=0$ and $\pi$. Here $J_K=6t_c$, and $T=0.004J_K$ in both cases.  } \label{fig6}
\end{figure}

The second case is a two-channel Kondo case $K=4S$ with a band of $3/4$-filled conduction electrons in Fig.\,\ref{fig6}. For an infinitesimal $t_b$ the low energy spectrum shows unit cell-doubling and consequently two pairs of low-energy spinon and holon modes. A stronger $t_b$ breaks the symmetry between these two pairs.

\subsection{4. Emergent dispersion {in 1D}}

According to the self-consistency equations in Eqs.\,\eqref{app:self-cons-1}--\eqref{app:self-cons-G}, if the input $G_b$ and $G_\chi$ are both local and $t_b=0$, then the updated Green's functions will remain local. Thus naively, without $t_b$ any Kondo lattice problem always reduces to a 0D impurity one. As shown in the Letter, 1D 2CKL is quite different from the impurity case when $t_b \neq 0$. Decreasing $t_b$ from large values will reduce the bandwidths of low energy spinon and holon modes, but below a certain $t_b^{*}$, these bandwidths will cease to decease. This is the emergent dispersion dynamically generated due to an amplified RKKY interaction, as illustrated in Fig.\,\ref{fig2}.

The emergent dispersion lowers the free energy of the system compared to the local solution, as seen in the entropy inset of Fig.\,\ref{fig2}(b). At high temperature, denoted by $T_h$, the 0D and infinitesimal-$t_b$ 1D$'$ state have the same entropy $S$ and free energy $F$. At low temperature, denoted by $T_l$, $S_{\mathrm{0D}} > S_{\mathrm{1D}'}$. Since $F(T_l) = F(T_h) + \int_{T_l}^{T_h}S\mathrm{d}T$, the solution of 1D$'$ gives a lower free energy. Therefore it is the genuine solution of the system. 

\begin{figure}
	\includegraphics[width=1\linewidth]{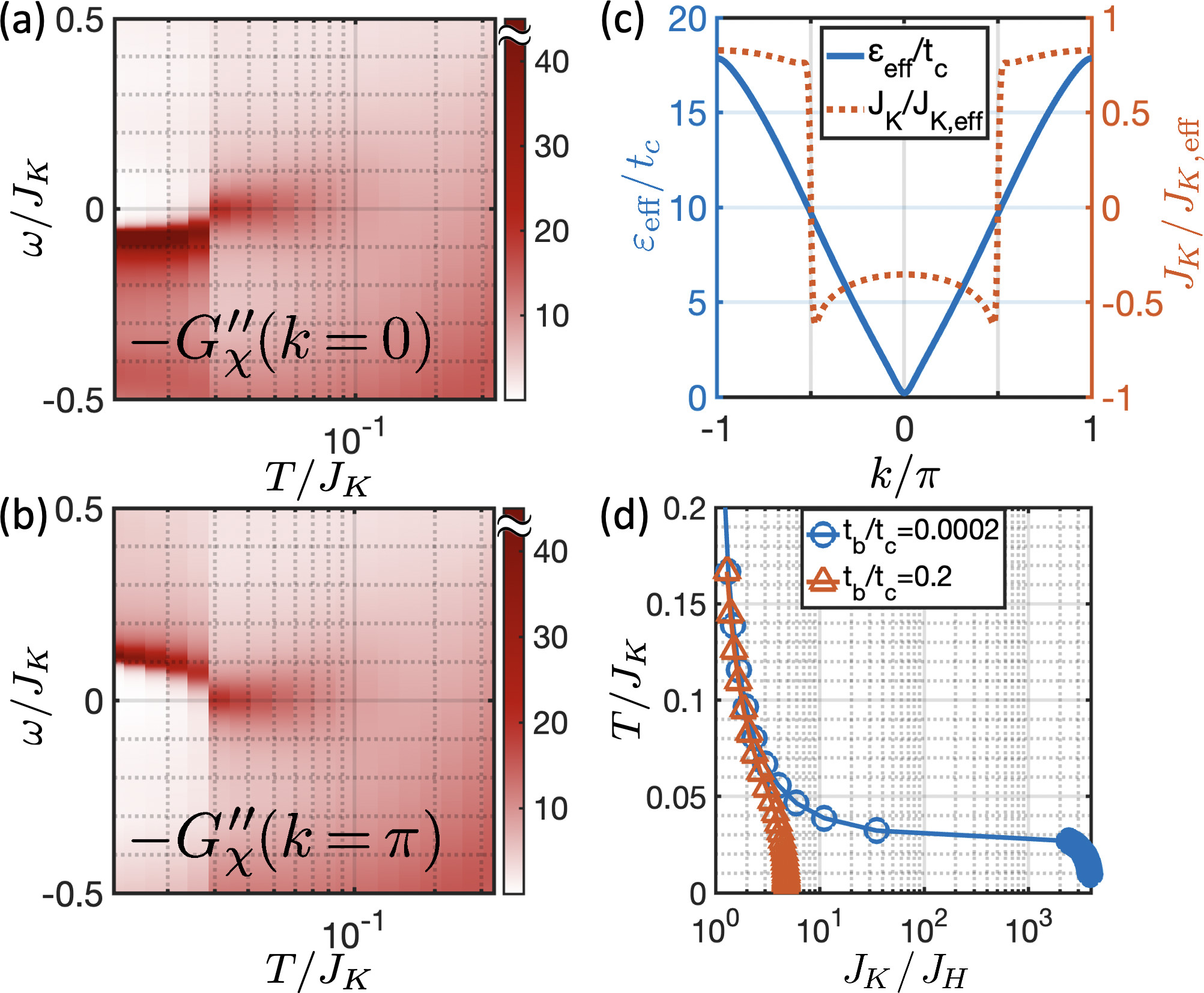}
	\caption{ (a--b) Without $J_H$, holon (and spinon) dispersion can emerge at low temperature in the presence of a tiny seed at the beginning of the self-consistency loop at each $T$. (c) The effective spinon energy and Kondo coupling for 1D 2CKL at $t_b=0.2t_c$. The low temperature behavior here (at $T=0.0044J_K$) is similar to the case with an infinitesimal $t_b$ in Fig.\,\ref{fig2}. (d) Effective Heisenberg coupling in 1D 2CKL for an infinitesimal and finite $t_b$.}\label{fig-app-JH}
\end{figure}

This phenomenon is best demonstrated in a seeding numerical experiment for the zero static hopping case, i.e., $t_b=0$. Before the self-consistency loop starts at each $T$, one can add a tiny $k$-dependent $\tilde{\Sigma}(k,\omega)T$ to the self-energy of spinons (or holons) used to construct the Green's function, which is at Step \ref{prog:solve-one-T} of the solver. The form of this seed ensures that it decreases with $T$. The $\omega$ dependence is nonessential. At high temperature, $\tilde{\Sigma}$ is rapidly suppressed by the self-consistency iterations and the Green's functions remain local, as seen in Fig.\,\ref{fig-app-JH}(a,b). Below a {critical} temperature, a part of the dispersion is exponentially amplified until it saturates. {Since the seed magnitude diminishes with $T$, this critical temperature must be finite and independent of $\| \tilde{\Sigma} \|$}.

Another way to observe this effect is to use an infinitesimal $t_b \ll t_b^{*}$. The emergent dispersion will then dominate the shape of low energy modes. {This is also manifested in $J^{-1}_{K,\text{eff}}$ and $\varepsilon\dn_{\text{eff}}$, shown in Fig.\,\ref{fig2}.} {Similar to the case with self-energy seeding, the temperature for emergent dispersion onset is finite as $t_b \to 0$.}  {Compared to the case with a finite $t_b$ in Fig.\,\ref{fig-app-JH}(c), they behave qualitatively the same. Therefore, at both finite and infinitesimal $t_b$ the system flow to the same fixed point.}

Without $t_b$, the Lagrangian of Eq.\,\eqref{eqLag} is invariant under simultaneous Galilean boosts of spinons and holons, that is $(b_j,\chi_j) \to e^{i k j}(b_j,\chi_j) $. Consequently, the emergent dispersing mode here can freely translate in $k$. In the numerics, the exact form of $\tilde{\Sigma}$ will determine the center of this dispersion. Different forms of $\tilde{\Sigma}$, including random ones and longer range hopping terms, e.g.\ $\cos(n k)$, all yield the same low energy spectrum up to momentum translations. Therefore, the emergent dispersion is robust. Without loss of generality, we set the center of emergent spinon dispersions to $k=0$.

Another manifestation of the emergent dispersion is the behavior of effective $J_H$ as the system cools down. In 1D, variational principle applied to $t_b$ gives \cite{Parcollet1997,Komijani18}
\begin{eqnarray}
	\hspace{-4em}J_H^{-1} \!& = &\! \frac{1}{L} \frac{1}{2 t_b}  \sum_j \sbraket{b_j^{\dagger} b_{j+1}^{\phantom{\dagger}}+\mathrm{H.c.}} \\
	\!& = &\! -\frac{1}{L}\sum_k\int \frac{\mathrm{d}\omega}{\pi t_b} n_B(\omega) \cos(k) G_b^{\prime\prime}(k,\omega+i\eta).
\end{eqnarray}
For a fixed $t_b>t_b^{*}$, Fig.\,\ref{fig-app-JH}(c) shows that $J_H$ gradually decreases with $T$ until it settles at a finite zero-temperature value. For $t_b \ll t_b^{*}$, we see that when the dispersion emerges, the effective $J_H$ has a steep drop to a value close to zero. This shows that no Heisenberg coupling is needed in 1D for the spinons to become mobile.
\subsection{5. Susceptibilities}
\subsubsection{Channel Susceptibility}
\begin{figure}[h!]
	\includegraphics[width=1\linewidth]{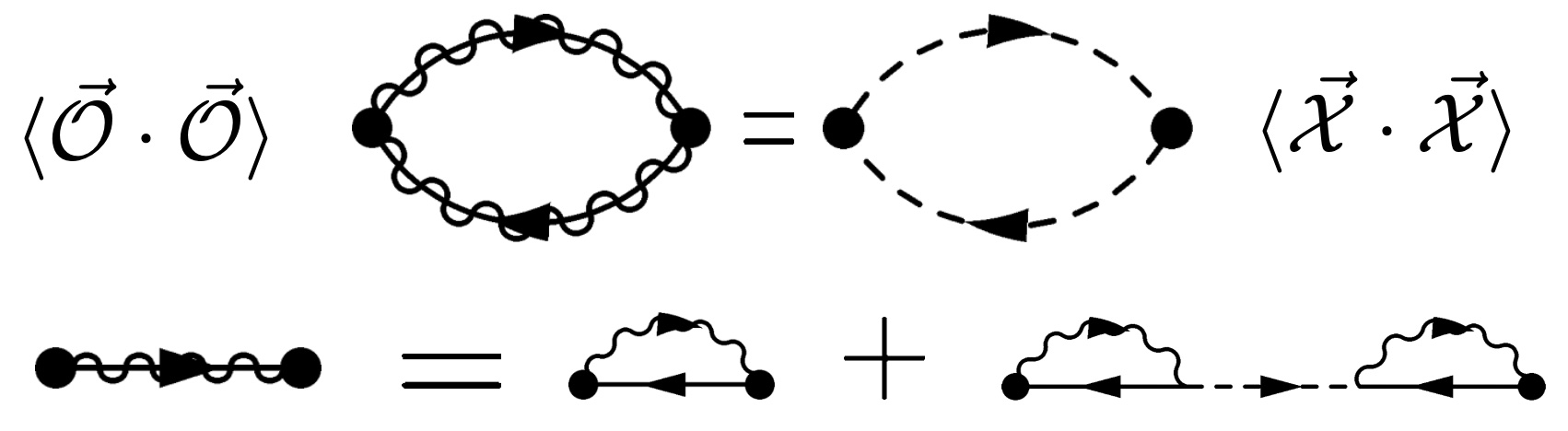}
	\caption{Equivalence of $\langle\vec{\cal O}\cdot\vec{\cal O}\rangle$ and $\langle\vec{\cal X}\cdot\vec{\cal X}\rangle$.}\label{figfeynman}
\end{figure}
In the presence of channel anisotropy, 
\begin{equation}
	H=H_c+\sum_jc\dg_{j,a\alpha}\Big(J_K\delta_{ab}+{\Delta \vec{J}_j\cdot\vec\tau_{ab}}\Big)c\dn_{j,b\beta}S_{j,\beta\alpha},
\end{equation}
the holon term in the Lagrangian is modified to
\begin{equation}
	\frac1{J_K}\sum_j\bar\chi_{j,a}\chi_{j,a} \ \to \ \frac{1}{J_K}\sum_j\bar\chi_{j,a}\Big(\bb 1+\frac{\Delta \vec{J}_j\cdot\vec\tau}{J_K}\Big)^{-1}_{ab}\chi_{j,b}.\nonumber
\end{equation}
Here $\tau^i$ with $i=1\dots K^2-1$ are generators of SU($K$) normalized according to ${\rm Tr}[\tau^i\tau^j]=\frac{1}{2}\delta^{ij}$. Expanding in small $\Delta J$, the Lagrangian is changed by
\begin{eqnarray}
	\Delta {\cal L}[\Delta J] &\!\!=\!\!& -\sum_j\bar{\chi}_{j,a}\frac{\Delta\vec{J}_j\cdot\vec\tau_{ab}}{J_K^2}\chi_{j,b}\nonumber\\
	& & +\sum_{j,ik}\bar{\chi}_{j,a}\frac{\Delta J_{j,i}\Delta J_{j,k}(\tau^i\tau^k)_{ab}}{J_K^3}\chi_{j,b},
\end{eqnarray}
up to $O(\Delta J^2)$. The channel susceptibility can be derived from taking derivatives of $\log Z$ w.r.t. $\Delta J$. We define
\begin{equation}
	\chi_{\mathrm{ch}} (\vec{r}) \equiv\sum_i \left. \frac{\partial^2  \ln Z + \partial^2 \ln \det [\beta (J_K + \Delta J^{ab})] }{\partial (\Delta J_{\vec{r}}^i) \partial (\Delta J_{\vec{0}}^i)} \right \rvert_{\Delta J=0}\hspace{-.5cm}
	\label{Xch-dF}
\end{equation}
where $\vec r\equiv(j,\tau)$. Here the second term comes from the free path integral $Z_\chi^0 = - \det [\beta (J_K + \Delta J)]$ of the Hubbard-Stratonovitch field $\chi$, which must be subtracted from the free energy. The first term gives
\begin{eqnarray}
	\left. \frac{\partial^2 \ln Z}{\partial (\Delta J_{\vec{r}}^i) \partial (\Delta J_{\vec{0}}^i)} \right \rvert_{\Delta J=0} \!& = & \frac{1}{J_K^4} \left [\tau_{a b}^i \tau_{cd}^i \langle (\bar{\chi}_a \chi_b)_{\vec{r}} (\bar{\chi}_c
	\chi_d)_{\vec{0}} \rangle_{\mathrm{lc}} \right. \nonumber\\
	& & \left. \  - 2 J_K \tau_{a c}^i \tau_{cb}^i \langle \bar{\chi}_a \chi_b \rangle \delta (\vec{r}) \right ],
\end{eqnarray}
where $\langle A(\vec{r}) B(\vec{t}) \rangle_{\mathrm{lc}}\equiv\langle A(\vec{r}) B(\vec{t}) \rangle-\langle A(\vec{r}) \rangle \langle B(\vec{t}) \rangle$ denoting the linked clusters. In the large-$N$ limit, $\langle (\bar{\chi}_a \chi_b)_{\vec{r}} (\bar{\chi}_c \chi_d)_{\vec{0}} \rangle_{\mathrm{lc}}=G_\chi(\vec{r})G_\chi(-\vec{r})\delta_{ad}\delta_{bc}$.
Noting that $g_\chi(\vec{r})=-J_K \delta(\vec{r})$, R.H.S.\ becomes
\begin{equation}
	J_K^{-4} [G_{\chi} (\vec{r}) G_{\chi} (- \vec{r}) + 2 G_{\chi}(\vec{r}) g_{\chi} (\vec{r})] \mathrm{Tr} \left [\tau^i \tau^i \right ].
	\label{dlnZdJJ}
\end{equation}
The second term in Eq.\,\eqref{Xch-dF} gives the same expression as Eq.\,\eqref{dlnZdJJ} with an opposite sign and all $G_\chi \to g_\chi$. Therefore, we find for the channel susceptibility
\begin{equation}
	\chi_{\mathrm{ch}} (\vec{r}) = J_K^{-4} [G_{\chi} (\vec{r}) -g_{\chi} (\vec{r})] [G_{\chi} (- \vec{r}) - g_{\chi} (- \vec{r})].
\end{equation}
This expresses $\chi_{\mathrm{ch}}$ using $\langle {\cal X X} \rangle$ correlators.

\begin{figure}
	\includegraphics[width=.95\linewidth]{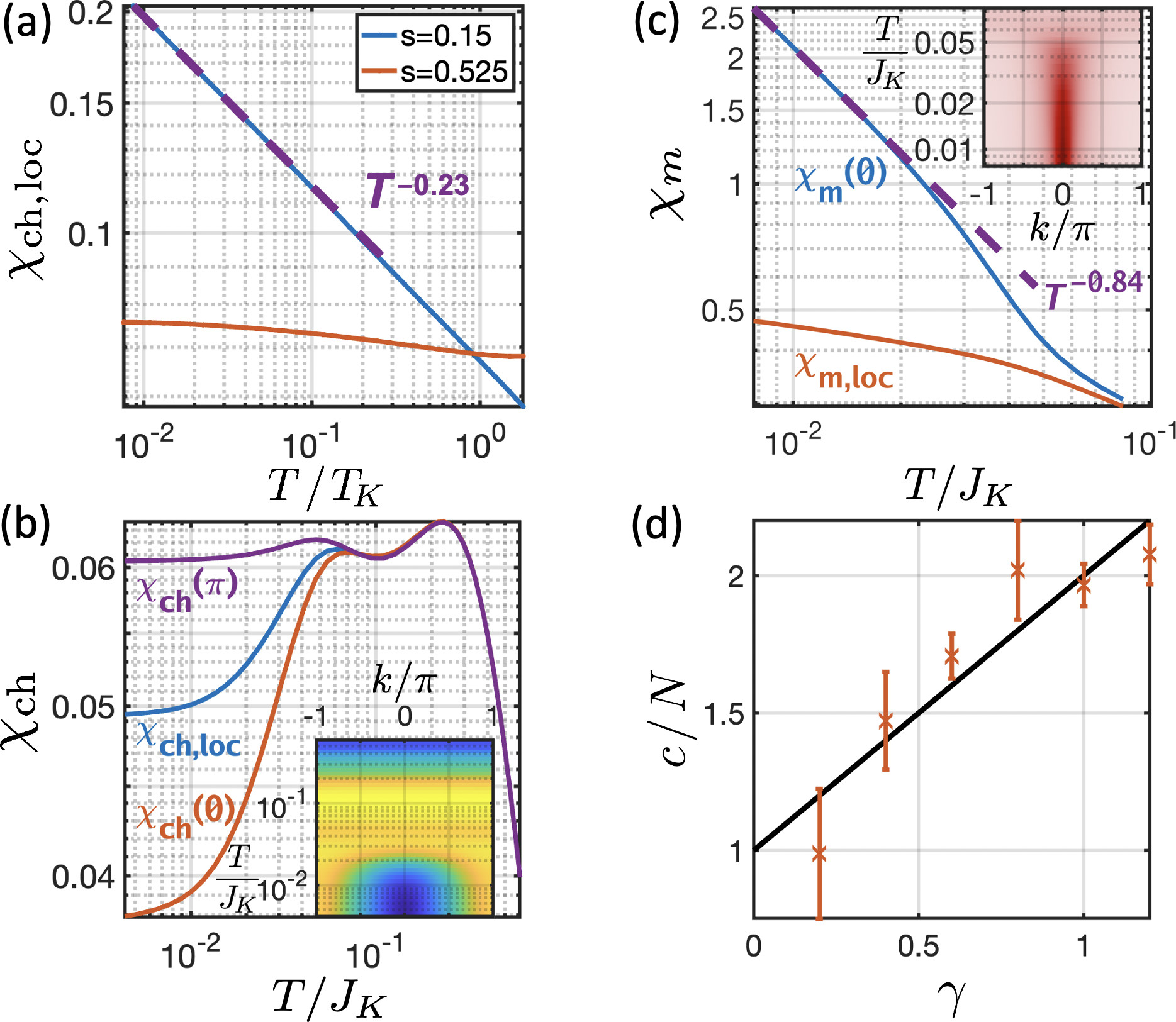}
	\caption{(a) Local static channel susceptibility of a double-screened 0D Kondo impurity, or a 1D chain without $J_H$. A diverging case at $s=0.15$, and a vanishing case at $s=0.525$ are shown. Power laws can only be extracted from diverging cases. (b) Static channel susceptibilities, $\chi_{\text{ch,loc}}$ and $\chi_{\text{ch}}(q)$, of an 1D 2CKL with $s=0.15$. They are vanishing as $T \to 0$ for all $s$. Inset shows the momentum resolved static $\chi_{\text{ch}}(q)$. The 1D case here has $J_K=6t_c$ and $t_b=0.2t_c$. (c) Uniform [$\chi_m(q=0)$] and local ($\chi_{m,\mathrm{loc}}$) static magnetic susceptibility of 1D 2CKL with $s=0.3$ and $\Delta_b=0.29$. Inset shows evolution of $\chi_m(q,\omega=0)$ with $T$. (d) The central charge $c$ extracted from heat capacity $C/T=(\pi k_B^2/6v)c$.}\label{fig:susCh}
\end{figure}



On the other hand, from the $\langle \vec{\cal O}\cdot\vec{\cal O}\rangle$ Feynman diagram in Fig.\,\ref{figfeynman} we have
\begin{equation}
	\sbraket{\vec{\cal O}(\vec r)\cdot\vec{\cal O}}=\delta G_\chi(\vec r)\delta G_\chi(-\vec r),
\end{equation}
where $\delta G_\chi$ is defined as
\begin{equation}
	\delta G_\chi(\vec r)\equiv \Sigma_\chi(\vec r)+\sum_{\vec r_1\vec r_2}\Sigma_\chi(\vec r-\vec r_1)G_\chi(\vec r_1-\vec r_2)\Sigma_\chi(\vec r_2).
\end{equation}
In momentum and frequency space
\begin{equation}
	\delta G_\chi(k,{\rm z})=\Sigma_\chi(1+G_\chi\Sigma_\chi)=\frac{1}{J_K^2}[G_\chi(k,{\rm z})+J_K].
\end{equation}
Thus, $\delta G_\chi = (G_\chi-g_\chi)/J_K^2$. Therefore, as Fig.\,\ref{figfeynman} suggests, the two approaches agree.


In Fig.\,\ref{fig:susCh}(a,b), we show the static channel susceptibilities of the 0D and 1D 2CKL. As discussed in the Letter, in all dimensions the local susceptibility is $\chi_{\mathrm{ch}}(x=0)=\int_0^\beta \mathrm{d}\tau \delta G_\chi(0,\tau) \delta G_\chi(0,-\tau) \sim T^{4\Delta_\chi-1}$, and in 1D the uniform susceptibility is $\chi_{\mathrm{ch}}^{\mathrm{1D}}(q=0) = \int \mathrm{d}^2 r \delta G_\chi(\vec r) \delta G_\chi(-\vec r) \sim T^{4\Delta_\chi-2}$. For 0D or $\infty\text{D}$, the $\chi_{\mathrm{ch}}(x=0)$ changes from diverging to vanishing at low temperature as $s$ increases, according to Eq.\,\eqref{eq0D}. For 1D, the static channel susceptibilities are always vanishing according to Eq.\,\eqref{expo1D}. Since the Green's functions we computed have both a scaling part and a regular part, only diverging scaling laws may be reliably extracted. The regular part will typically overwhelm the vanishing components, except at very low (zero) frequencies as is the case for $v_\chi^{\mathrm{FS}}$ [Fig.\,\ref{fig4}(a)].

\subsubsection{Magnetic susceptibility}

We show the magnetic susceptibilities in 1D for $s=0.3$ in Fig.\,\ref{fig:susCh}(c). Here the uniform static magnetic susceptibility $\chi_m(q=0)$ is diverging, revealing a magnetic instability as discussed in the Letter. The local static susceptibility $\chi_m(x=0)$ is vanishing in this case, but will diverge for $s>0.375$. The inset shows $\chi_m(q)$ vs.\ $T$. At low $T$, $\chi_m(q)$ becomes sharply peaked at $q=0$.

\subsection{{6. Dispersion and ground state entropy in $\infty$D}}

\begin{figure}
	\includegraphics[width=0.8\linewidth]{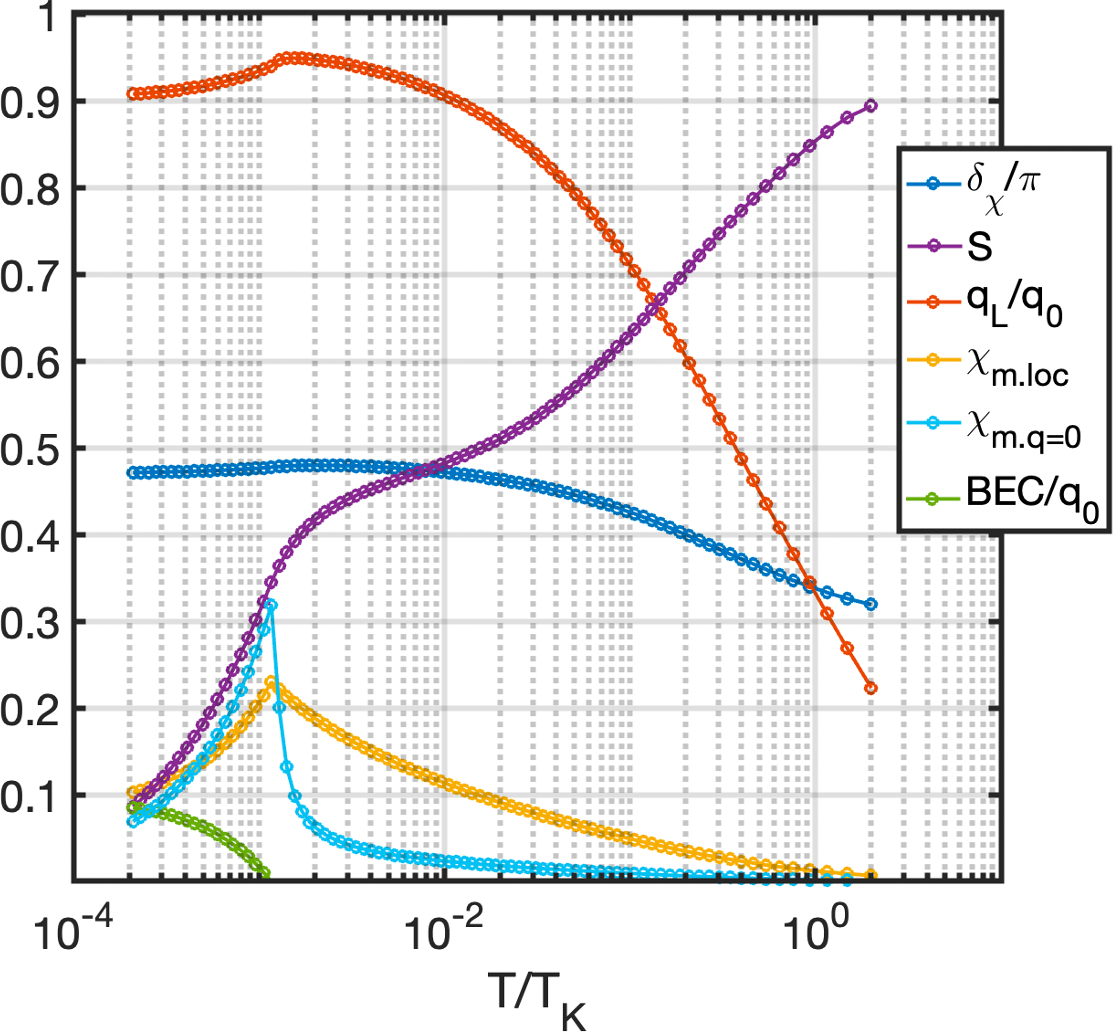}
	\caption{{The temperature dependence of various parameters in a double-screened ($K{\!}=\!4S$) Kondo lattice in $\infty$D ($z=\infty$ Bethe lattice). The parameters shown are phase shift $\delta_\chi/\pi$, entropy $S$, confined spinon fraction $q_L/q_0$, local $\chi_{m.loc}$ and uniform $\chi_{m.q=0}$ magnetic susceptibilities, as well as the condensate fraction $\mathrm{BEC}/q_0$ which gives the magnetization using Schwinger bosons. Note that the entropy goes to zero only when the system starts to order magnetically. The setup parameters are $s=0.35$, $J/t_c=1.88$, and $t_b/t_c=0.0024$. }}\label{fig:ooDentropy}
\end{figure}

{In the $z=\infty$ Bethe lattice, we find no emergent dispersion generated by the self-consistency solver. This comes from taking the infinite dimension limit first, which is a singular limit. Therefore, the dispersion is not a valid mechanism in this case to eliminate the extensive ground state entropy. However, the system is prone to other forms of symmetry breaking, such as spin or channel magnetization, and the entropy will then decrease to zero. An example is shown in Fig.\,\ref{fig:ooDentropy}.}

\subsection{7. Scaling analysis for impurity problem}
For the sake of completeness, we review the scaling analysis for the multichannel Kondo problem using Schwinger bosons \cite{Parcollet1997}, which also applies to the infinite-coordination lattice problem we have studied in this Letter. For the spinon and holon we use the $T=0$ ansatz
\begin{equation}
	G_b(\tau)=b_b\frac{1}{\abs{\tau}^{2\Delta_b}},\quad
	G_\chi(\tau)=b_\chi\frac{\sgn\tau}{\abs{\tau}^{2\Delta_\chi}}.
\end{equation}
Using the self-energy formulae
\begin{equation}
	\Sigma_b(\tau)=-\gamma g_c(\tau)G_\chi(\tau),\quad \Sigma_\chi(\tau)=g_c(-\tau)G_b(\tau),
\end{equation}
and that $g_c(\tau)=-{\rho_c}/{\tau}$, we have 
\begin{equation}
	{
		\Sigma_b(\tau)=-b_\chi\gamma\rho_c\frac{1}{\abs{\tau}^{2\Delta_{\chi}\!+\!1}},
		\ \Sigma_\chi(\tau)=b_b\rho_c\frac{\sgn\tau}{\abs{\tau}^{2\Delta_b+1}}.
	}
\end{equation}
Fourier transform of the Green's function are
\begin{eqnarray}
	G_b(i\omega)&=&2b_b\abs{\omega}^{2\Delta_b-1}\Gamma(1-2\Delta_b)\sin{\pi}\Delta_b,\\
	G_\chi(i\omega)&=&2b_\chi\abs{\omega}^{2\Delta_\chi-1}\Gamma(1-2\Delta_\chi)\cos{\pi}\Delta_\chi\sgn(\omega).\nonumber
\end{eqnarray}
with similar expressions for $\Sigma_b$ and $\Sigma_\chi$.  $G_c$ can be computed from $G_\chi$ in the $\Delta_\chi\to1/2$ limit.  The zero and the pole of $G_c$ cancel each other in this limit. Next we plug these into the Dyson equations:
\begin{equation}
	G_b(i\omega)[i\omega-\lambda-\Sigma_b(i\omega)]=1,\quad\! G_\chi(i\omega)[-J_K^{-1}-\Sigma_\chi(i\omega)]=1.\nonumber
\end{equation}
In the scaling limit, the numerical solutions show that $G_\chi\Sigma_\chi=G_b\Sigma_b=-1$. For this equation to hold, the powers of frequency have to cancel from the left side, i.e. $2\Delta_b+2\Delta_\chi=1$, and the amplitudes must match, i.e.
\begin{eqnarray}
	\hspace{-1.5em}4\rho \gamma b_bb_\chi\sin^2(\pi\Delta_b)\Gamma(1-2\Delta_b)\Gamma(2\Delta_b-1)&\!=\!&-1, \nonumber\\
	4\rho b_\chi b_b\sin^2(\pi\Delta_b)\Gamma(2\Delta_b)\Gamma(-2\Delta_b)&\!=\!&-1.
\end{eqnarray}
Using $\Gamma(z+1)=z\Gamma(z)$, the ratio finally gives 
\begin{equation}
	2\Delta_b=\frac{1}{1+\gamma}, \qquad 2\Delta_\chi=\frac{\gamma}{\gamma+1}.
\end{equation}
For the case of $\infty$D Bethe lattice, we can write
\begin{equation}
	G_x^{-1}=\Omega\dn_x-t_x^2G\dn_x,
\end{equation}
for $x=b,c$ where $\Omega_c({\rm z})={\rm z}+\mu_c$ but $\Omega_b({\rm z})={\rm z}+\mu_b-\Sigma_b({\rm z})$. The solution is
\begin{equation}
	t_xG_x=\Omega_x/2t_x-\sgn(\Omega'_x)\sqrt{(\Omega_x/2t_x)^2-1}.
\end{equation}
We are interested in the $t_b\to0$ limit of this expression. We find not surprisingly that
\begin{equation}
	t_bG_b=(\Omega_b/2t_b)[1-\sqrt{1+(2t_b/\Omega_b)^2}]\approx t_b/\Omega_x,
\end{equation}
showing that in this limit the lattice is dominated by the local impurity solution and so, exponents are the same.

\subsection{8. Ground state entropy of the impurity model}
The ground state entropy of $\frac{1}{2}\log 2$ discussed in the introduction is specific to the spin-1/2 SU(2) two-channel Kondo model, i.e., $N=K=4S=2$. For generic $(N,K,S)$, the result from Bethe ansatz and conformal field theory is \cite{Parcollet1997}
\begin{equation}
	S_{\rm imp}=\log\prod_{n=1}^{2S}\frac{\sin[\pi(N+n-1)/(N+K)]}{\sin[\pi n/(N+n)]}.
\end{equation}
In the large-$N,K,S$ limit (keeping $\gamma=K/N$ and $s=S/N$ constant), this gives
\begin{equation}
	S_{\rm imp}/N=\frac{1+\gamma}{\pi}[f_\gamma(1+2s)-f_\gamma(1)-f_\gamma(2s)],
\end{equation}
where $f_\gamma(x)=\int_0^{\pi x/(1+\gamma)}\log\sin(u)\,du.$
This fractional entropy is universally characterized by $(\gamma,s)$. It is fully consistent with the result of the large-$N$ theory as shown in Refs.~\onlinecite{Parcollet1997,Komijani18}. Specifically, it agrees with Fig.~\ref{fig1}(c).

\subsection{9. Details of scaling analysis in 1D case}
This section follows closely the impurity solution. We use the symbol $b_j(\tau)\sim \sqrt {\rm a}b(x,\tau)$ for $x=j{\rm a}$ to refer to the low-momentum $k\sim 0$ content of the spinons. We also expand fermionic field around the Fermi energy:
\begin{eqnarray}
	{\rm a}^{-1/2}c_j(\tau)&\sim& e^{ik_Fx}c_R(x,\tau)+e^{-ik_Fx}c_L(x,\tau),\\
	\chi_j(\tau)&\sim& e^{ik_Fx}\chi_R(x,\tau)+e^{-ik_Fx}\chi_L(x,\tau).
\end{eqnarray}
The interaction term becomes
\begin{equation}
	H_{\mathrm{int}}=\frac{1}{\sqrt N}\sum_{a\alpha}\int{dx}[(\chi\dg_{Ra}c\dg_{La\alpha}+\chi\dg_{La}c\dg_{Ra\alpha})b\dn_{\alpha}+\mathrm{H.c.}].\label{eqHint}
\end{equation}
This leads to the self-energies
\begin{eqnarray}
	\Sigma_{\chi R/L}(\vec r)&=&G_{cL/R}(-\vec r)G_{b}(\vec r),\\
	\Sigma_b(\vec r)&=&-\gamma[G_{cR}(\vec r)G_{\chi L}(\vec r)+G_{cL}(\vec r)G_{\chi R}(\vec r)].\nonumber
\end{eqnarray}
In terms of these, the lattice Green's functions are
\begin{eqnarray}
	{\rm a}^{-1}\breve g_c(x,\tau)&\!\!=\!\!&e^{ik_Fx}g_{R}(x,\tau)\!+\!e^{-ik_Fx}g_{L}(x,\tau),\\
	\breve G_\chi(x,\tau)&\!\!=\!\!&e^{ik_Fx}G_{\chi R}(x,\tau)\!+\!e^{-ik_Fx}G_{\chi L}(x,\tau),\quad\\
	{\rm a}^{-1}\breve G_b(x,\tau)&\!\!=\!\!&G_b(x,\tau).
\end{eqnarray}
The ansatzes put forward in the Letter are
\begin{equation}
	G_b=-\bar\rho\Big(\frac{1}{\bar zz}\Big)^{\Delta_b}\hspace{-.25cm}, \quad G_{\chi R/L}=\frac{-1}{2\pi}\Big(\frac{1}{\bar z}\Big)^{\Delta_\chi\pm \frac12}\Big(\frac{1}{z}\Big)^{\Delta_\chi\mp \frac12}.\hspace{-.15cm}\nonumber
\end{equation}
The Fourier transform of the imaginary-time function is straightforward:
\begin{equation}
	\hspace{-0.6em}G_b(k,i\omega)\!=\!-\bar\rho\!\intinf\!{\mathrm{d}x}\!\intinf\!{\mathrm{d}\tau}e^{-i(kx-\omega\tau)}\Big(\frac{1}{\bar zz}\Big)^{\Delta_b}.
\end{equation}
Defining $z\equiv u\tau+ix=re^{i\phi}$ and $q\equiv k+i\omega/u=\abs{q}e^{i\theta}$,
\begin{equation}
	kx-\omega\tau=\im{z\bar q}=r\abs{q}\sin(\phi-\theta)
	,\label{eqpolar}
\end{equation}
and we have
\begin{eqnarray}
	G_b(k,i\omega)=-2\pi\bar\rho u_b^{-1}\!\int{r\mathrm{d}r}\!\int_0^{2\pi}\frac{d\phi}{2\pi}e^{-i\abs{q}r\sin(\phi-\theta)}\frac{1}{r^{2\Delta_b}}.\nonumber
\end{eqnarray}
This integral is the $n=0$ version of a more general integral that we will encounter again. So, let us define
\begin{eqnarray}
	{\cal I}_{n,\Delta}(q)&\equiv&\int\frac{r\mathrm{d}r}{r^{2\Delta}}\int\frac{{\mathrm{d}\phi}}{2\pi}e^{-i[r\abs{q}\sin(\phi-\theta)+n\phi]}\nonumber\\
	&=&\abs{q}^{2(\Delta-1)}(-1)^ne^{-in\theta}\zeta_n(\Delta),
\end{eqnarray}
where we have used a $\phi\to \phi+\theta-\pi/2$ shift to write it in terms of the $\zeta$ function
\begin{equation}
	\zeta_n(\Delta)\equiv\intoinf{\mathrm{d}x}x^{1-2\Delta}R_n(x), 
\end{equation}
which by itself is written in terms of 
\begin{eqnarray*}
	R_n(x)&\equiv& (-i)^{n}\int_0^{2\pi}\frac{\mathrm{d}\phi}{2\pi} e^{-in\phi}e^{ix\cos\phi}\nonumber\\
	&=&(-i)^{n}\int_0^{2\pi}\frac{\mathrm{d}\phi}{2\pi} e^{-in\phi}\sum_mi^{m}J_m(x)e^{im\phi}=J_n(x).
\end{eqnarray*}
Therefore, zeta-function is nothing but the \emph{Mellin transform} of the Bessel function:
\begin{equation}
	\zeta_{n\ge 0}(\Delta)={2^{1-2\Delta}}\frac{\Gamma(1-\Delta+n/2)}{\Gamma(\Delta+n/2)},\label{eqzeta}
\end{equation}
valid for $-3/2<2\Delta-2<n$. Since $J_{-n}(x)=(-1)^nJ_n(x)$ we can express $\zeta_{-n}(\Delta)=(-1)^n\zeta_{n>0}(\Delta)$ in terms of $\zeta_{n\ge 0}$. Using the ${\cal I}_{0,\Delta}(q)$ integral we readily find
\begin{equation}
	G_b(k,i\omega)=-2\pi\bar\rho u_b^{-1}\abs{q}^{2(\Delta_b-1)}\zeta_0(\Delta_b),
\end{equation}
where $\zeta_0(\Delta)=2^{1-2\Delta}{\Gamma(1-\Delta)}/{\Gamma(\Delta)}$ from Eq.\,\pref{eqzeta}. Analytical continuation $i\omega\to\omega+i\eta$ leads to
\begin{eqnarray}
	\!\!\!\!\!\!\! G''_b(k,\omega+i\eta)&\!\!\!=\!\!\!&{\bar\rho u_b^{-1}\!\abs{\frac{\omega^2\!-\! u^2k^2}{4u}}^{\Delta_b-1}\!\Gamma^2(1-\Delta_b)}\\
	&&\hspace{-5em}\times\frac{1}{2}\sin^2\pi\Delta_b\sin\Big[\sgn(\omega/u+k)+\sgn(\omega/u-k)\Big].\nonumber
\end{eqnarray}
\sepline
For the holons we have
\begin{eqnarray}
	G_{\chi R}(k,i\omega)&=&-\frac{u_\chi^{-1}}{2\pi}\intinf\!{\mathrm{d}x}\\
	&&{\hspace{-4em}\intinf\!{\mathrm{d}\tau}e^{-i(kx-\omega\tau)}\Big({\frac{1}{x^2+u^2\tau^2}}\Big)^{\Delta_\chi}e^{+i\angle(u\tau+ix)}.}\nonumber
\end{eqnarray}
This is the $n=-1$ case of ${\cal I}_{n,\Delta_chi}(q)$ integral. Therefore,
\begin{equation}
	{\!\! G_{\chi R}(k,i\omega)=-u_\chi^{-1} (\bar qq)^{(\Delta_\chi-1)}\frac{\abs{q}}{\bar q}\zeta_{1}(\Delta_\chi),}
\end{equation}
and similarly for left-movers
\begin{equation}
	{G_{\chi L}(k,i\omega)=u_\chi^{-1} (\bar q q)^{(\Delta_\chi-1)}\frac{\abs{q}}{q}\zeta_{1}(\Delta_\chi).}
\end{equation}
We use these to find the Fourier transform of lattice Green's functions. Defining shifted variables
\begin{eqnarray*}
	q_n\equiv(k-nk_F)+i\omega/u,\quad\theta_n\equiv\angle[(k-nk_F)+i\omega/u]
\end{eqnarray*}
and using $\zeta_1(1/2)=1$, 
the lattice Green's functions are
\begin{eqnarray}
	{\rm a}^{-1}\breve G_b(k,i\omega)&=&-2\pi\frac{\bar\rho }{u_b}\zeta_0(\Delta_b)\abs{q}^{2\Delta_b-2}\\
	{\rm a}^{-1}\breve g(k,i\omega)&=&\frac{1}{v_F}\Big[\frac{-1}{\bar q_1}\Big]+\frac{1}{v_F}\Big[\frac{1}{q_{-1}}\Big]\\
	\breve G_\chi(k,i\omega)&=&\frac{1}{u_\chi}\zeta_1(\Delta_x)\Big\{-\abs{q_1}^{2(\Delta_\chi-2)}\frac{\abs{q_1}}{\bar q_1}\\
	&&\hspace{5.5em}+\abs{q_{-1}}^{2(\Delta_\chi-2)}\frac{\abs{q_{-1}}}{q_{-1}}\Big\}\nonumber.
\end{eqnarray}

Assuming that the bare Green's functions computed above are the exact ones, the boson self-energy is
\begin{equation}
	\Sigma_b(x,\tau)=-\frac{\gamma}{2\pi^2}\Big(\frac{1}{\bar zz}\Big)^{\Delta_\chi+1/2},
\end{equation}
and the holon self-energy is
\begin{equation}
	\Sigma_{\chi R/L}(x,\tau)=-\frac{\bar\rho}{2\pi}\Big(\frac{1}{\bar z z}\Big)^{\Delta_b+1/2}e^{\pm i\angle z}.
\end{equation}
To have conformal invariance, we have assumed all the velocities are the same $u_b=u_\chi=u_c=u$, and holons and conduction electrons have the same Fermi wavevectors. The Fourier transform can be computed as before:
\begin{eqnarray}
	\Sigma_b(k,i\omega)&=&-\gamma \frac{1}{\pi u}\zeta_0(\Delta_\chi\!+\!1/2)\!\abs{q}^{2\Delta_\chi-1}\qquad\\
	\Sigma_{\chi R/L}(k,i\omega)&=&\mp\frac{\bar\rho}{u}\zeta_1(\Delta_{b}\!+\!1/2)e^{\pm i\theta}\abs{q}^{2\Delta_b-1}.
\end{eqnarray}
For the self-consistency, we have to note that $G\Sigma\propto {\rm a}^{1-2\xi}$ and we must choose $\xi=1/2$. To first approximation, we neglect the $2k_F$ contributions. The self-consistency, then leads to [Note that in the second line, the signs do not match for $\Delta_\chi\in(0,0.5)$, therefore $\Delta_\chi\not\in(0,0.5)$.]
\begin{eqnarray}
	G_b\Delta\Sigma_b=-1 \ \to \ 2\gamma A\Big[\zeta_0(\Delta_b)\zeta_0(\Delta_\chi+1/2)\Big]\!\!&=&\!\!-1,\nonumber\\
	G_{\chi R/L}\Delta\Sigma_{\chi R/L}=-1 \ \to\  A\Big[\zeta_1(\Delta_b+1/2)\zeta_1(\Delta_\chi)\Big]\!\!&=&\!\!1,\nonumber
\end{eqnarray}
where we defined $A={\bar\rho }/{u_\chi u_b}$. Eliminating $A$ between these equations, and using $\Delta_b=3/2-\Delta_\chi$ we find
\begin{equation}
	2\gamma=-\frac{\zeta_1(2-\Delta_\chi)\zeta_1(\Delta_\chi)}{\zeta_0(3/2-\Delta_\chi)\zeta_0(\Delta_\chi+1/2)}.
\end{equation}
To solve this equation, we use the relation $\Gamma(\Delta)\Gamma(1-\Delta)={\pi}/{\sin\pi \Delta} $, the $\zeta$ function can be written as
\begin{equation}
	\zeta_n(\Delta)=2^{1-2\Delta}\frac{\sin\pi(n/2+\Delta)}{\pi}\Gamma(1-\Delta+n/2)\Gamma(1-\Delta-n/2).\nonumber
\end{equation}
This equation, together with $\Delta_1+\Delta_2=2$ and $\sin(\pi z)\Gamma(z)\Gamma(-z)=-\pi/z$ can be used to prove that 
\begin{equation}
	\zeta_n(\Delta_1)\zeta_n(\Delta_2)=\frac{-1/4}{(1-\Delta_1)^2-n^2/4},
\end{equation}
for $n=0,1$. Using these we find the solution
\begin{equation}
	2\gamma=-\frac{(\Delta_\chi-1/2)^2}{(\Delta_\chi-1)^2-1/4},
\end{equation}
which leads to
\begin{equation}
	\Delta_\chi=\frac{1+6\gamma}{2(1+2\gamma)}, \qquad \Delta_b=\frac{3}{2}-\Delta_\chi=\frac{2}{2(1+2\gamma)}.\label{app:expo1D}
\end{equation}

\begin{figure}[h!]
	\includegraphics[width=0.99\linewidth]{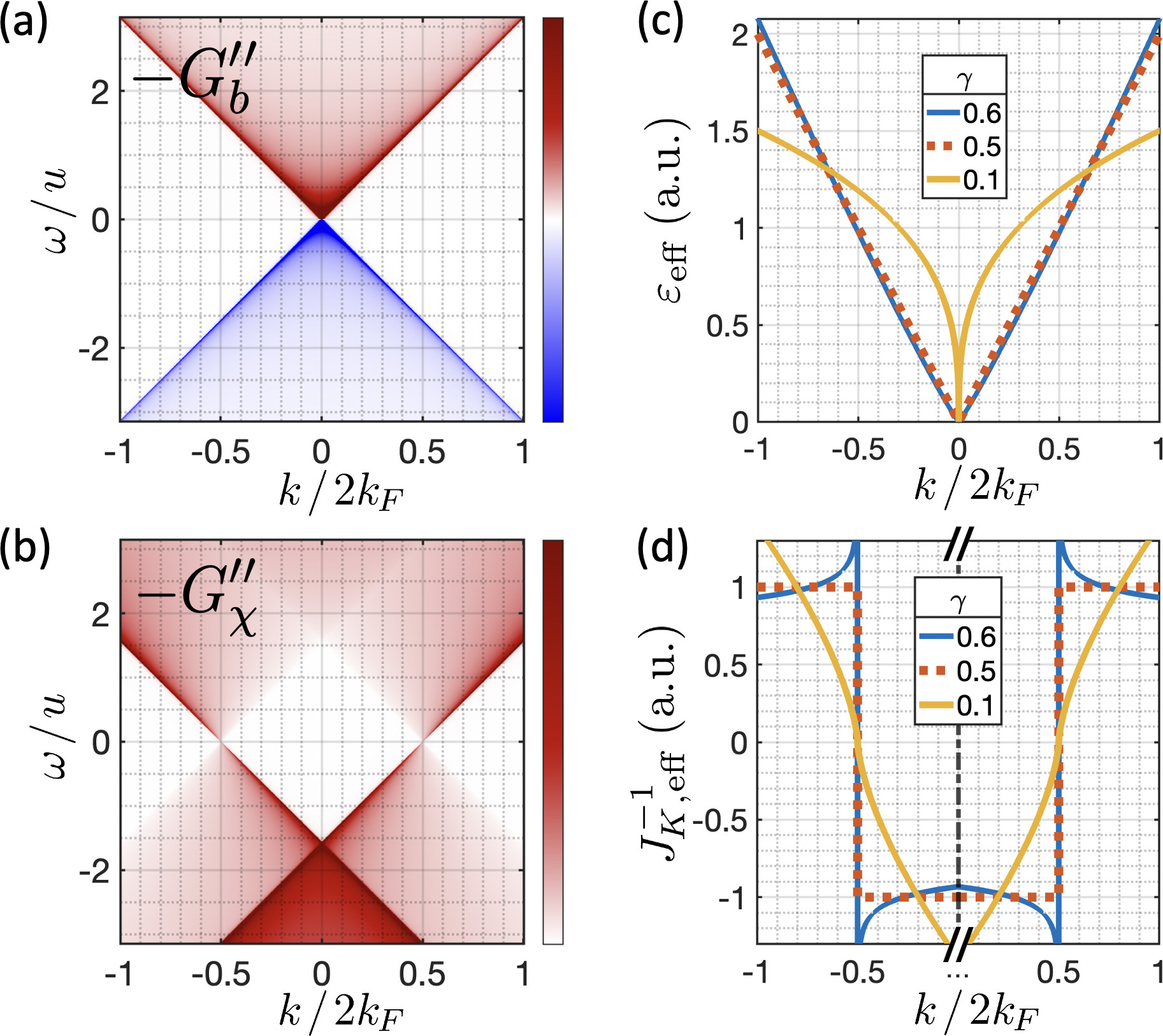}
	\caption{{The conformal solutions to large-$N$ 1D 2CKL derived here. The spectral weights of (a) spinons and (b) holons for $s=0.15$ ($\gamma=0.6$) capture the low frequency features in Fig.~\ref{fig3}. The effective dispersions of (c) spinons and (d) holons for several choices of $\gamma\equiv4s$ are plotted. Since the conformal solutions only hold for small $k$ and $\omega$, in Panel (d) we show only one of $G_{\chi L/R}$ close to left/right fermi surface ($k_F=\pm\pi/2$) at each side. The $\gamma=0.6$ lines resemble Fig.~\ref{fig2}.}}\label{fig:ansatz}
\end{figure}

{The conformal solutions to the large-$N$ 1D two-channel Kondo lattice we just derived are plotted in Fig.~\ref{fig:ansatz}. Note that the singularities at $\pm k_F$ in the effective holon dispersion $J^{-1}_{K,\mathrm{eff}}$ is physically allowed since the holon is incoherent there. This ``jump'' diminishes as $\gamma \to 0$.}

\vspace{.5cm}
\subsection{10. Large-$N$ limit of the Andrei-Orignac coset theory}
According to Ref.\ \onlinecite{Andrei2000}, using non-Abelian bosonization in the limit of large Heisenberg coupling between spins and away from any charge commensurate filling, the Hamiltonian can be written in the form of $\vec J_L\cdot\vec S_R+\vec J_R\cdot\vec S_L$ where $\vec S_{R/L}$ are SU$_1$(2) right/left-mover currents describing local moments and $\vec J_{R/L}$ are SU$_K$(2) right/left-mover currents describing the spin sector of fermions. Ref.\,\onlinecite{Andrei2000} shows that this system flows to the so-called chirally-stabilized fixed point, described by the coset theory,
\begin{eqnarray}
	c_{\rm coset}(2,K)&\equiv& c\Big[\frac{\mathrm{SU}_{K-1}(2)\times \mathrm{SU}_1(2)}{\mathrm{SU}_{K}(2)}\Big]\\
	&=&1-\frac{6}{(K+1)(K+2)}=0, \muad\frac{1}{2}, \muad \frac{7}{10}, \muad \frac{4}{5},\muad\cdots.\nonumber
	\label{eqOA}
\end{eqnarray}
For $K=2$ a dispersing Majorana model is predicted. Assuming that $\vec J_{R/L}$ and $\vec S_{R/L}$ are currents of SU($N$) WZW models, we propose a generalization of this theory
\begin{eqnarray}
	c_{\rm coset}(N,K)&\!\!\equiv\!\!& c\,\Big[\frac{\mathrm{SU}_{K-1}(N)\otimes \mathrm{SU}_1(N)}{\mathrm{SU}_{K}(N)}\Big]\\
	&\!\!=\!\!&(N^2\!-\!1)\Big[\frac{K-1}{N+K-1}+\frac{1}{N+1}-\frac{K}{N+K}\Big].\nonumber
\end{eqnarray}
In the limit of large-$N$ we find
\begin{equation}
	\lim_{N\to\infty}\frac{1}{N}c_{\rm coset}(N,\gamma N)=1-\frac{1}{(1+\gamma)^2}.
\end{equation}
However, in addition to this non-magnetic part, there is also a residual magnetic contribution, which is given by
\bea
c[SU_{K-1}(N)]&=&\frac{(N^2-1)(K-1)}{N+K-1}\to 
\frac{N^2\gamma}{1+\gamma}-\frac{N\gamma}{(1+\gamma)^2}\nonumber
\eea
The $N^2$ extensive part of this, together with the decoupled charge ($c=1$) and channel contribution
\be
c[SU_N(K)]=\frac{(K^2-1)N}{N+K}\to N^2\frac{\gamma^2}{1+\gamma}
\ee
gives the $c=N^2\gamma=NK$ of conduction electrons. The remaining O(N) part of the magnetic modes adds to the non-magnetic coset part to give the IR central charge
\be
c_{AO}=\Delta c[SU_{K-1}(N)]+c_{\rm coset}(N,K)=N\frac{\gamma}{1+\gamma}.
\ee

\end{document}